\documentclass[prb,aps,twocolumn]{revtex4}
\usepackage{amsmath}
\usepackage{amssymb}
\usepackage {graphicx}

\newcommand{\be}{\begin{equation}}
\newcommand{\ee}{\end{equation}}
\newcommand{\beq}{\begin{equation}}
\newcommand{\eeq}{\end{equation}}
\newcommand{\bea}{\begin{eqnarray}}
\newcommand{\eea}{\end{eqnarray}}
\newcommand{\w}{{\omega}}

\begin{document}

\title{Optical Integral and Sum Rule Violation}
\author{Saurabh Maiti, Andrey V. Chubukov}
\affiliation {~Department of Physics, University of Wisconsin,
Madison, Wisconsin 53706, USA}
\date{\today}


\begin{abstract}
The purpose of this work is to investigate the role of the
lattice in the optical Kubo sum rule in the cuprates.
We compute conductivities, optical integrals $W$,
 and  $\Delta W$  between superconducting and normal states  for
2-D systems with lattice dispersion typical of the cuprates for
four different models -- a dirty BCS model, a single Einstein
boson model, a marginal Fermi liquid model, and a collective
boson model with a feedback from superconductivity on a collective
boson.  The goal of the paper is two-fold. First, we analyze the
dependence of $W$ on the upper cut-off ($\omega_c$) placed on the
optical integral because in experiments $W$ is measured up to
frequencies of order bandwidth. For a BCS model, the Kubo sum rule
is almost fully reproduced at  $\omega_c$ equal to the bandwidth.
But for other models only ~$70\%$-$80\%$ of Kubo sum rule is
obtained up to this scale and even less so for $\Delta W$,
implying that the Kubo sum rule has to be applied with caution.
Second, we analyze the sign of $\Delta W$. In all models we
studied $\Delta W$ is positive
 at small $\omega_c$, then crosses zero and  approaches a
 {\it negative} value at large $\omega_c$, i.e. the optical integral in a
 superconductor is smaller than in a normal state.
The point of  zero crossing, however, increases with the interaction strength
 and in a collective boson model becomes comparable to the bandwidth
at strong coupling. We argue that this model exhibits the behavior consistent
 with that in the cuprates.
\end{abstract}

\maketitle

%%%%%%%%%%%%%%%%%%%%%%%%%%%%%%%%%%%%%%%%%%%%%%%%%%%%%%%%%%%%%%%%%%%%%%
\section{Introduction}

The analysis of sum rules for optical conductivity has a long
history. Kubo, in an extensive paper\cite{bib:Kubo Sum rule} in
1957, used a general formalism of a statistical theory of
irreversible processes to investigate the behavior of the conductivity
 in electronic systems. For a system of interacting
electrons, he derived  the expression for the integral of the real
part of a (complex) electric conductivity
$\sigma (\Omega)$ and found that it is independent on
 the nature of the interactions and reduces to
\begin{equation}
\int^{\infty}_{0} \, Re \, \sigma(\Omega) \, d\Omega =
\frac{\pi}{2}\frac{ne^2}{m} \label{n_1}
\end{equation}
Here  $n$ is the density of the electrons in the system and $m$ is
the bare mass of the electron.  This expression is exact provided
that the integration extends truly up to infinity, and its
derivation uses the obvious fact that at energies higher than the
total bandwidth of a solid, electrons behave as free particles.

The independence of the r.h.s. of Eq. (\ref{n_1}) on temperature
and the state of a solid (e.g., a normal or a superconducting
state -- henceforth referred to as NS and SCS respectively)
implies that, while the functional form of $\sigma (\Omega)$
changes with, e.g., temperature, the total spectral weight is
conserved and only gets redistributed between different
frequencies as temperature changes. This conservation of the total
weight of $\sigma (\Omega)$ is generally called a sum rule.

One particular case, studied in detail for conventional
superconductors, is the redistribution of the spectral weight
between normal and superconducting states. This is known as
Ferrel-Glover-Tinkham (FGT) sum rule:\cite{bib:FGT1, bib:FGT2}
\begin{equation}
\int^{\infty }_{0+} \,Re\,\sigma_{NS}(\Omega) = \int^{\infty
}_{0+} \,Re\, \sigma_{sc}(\Omega) + \frac{\pi n_s e^2}{2m}
\label{n_2}
\end{equation}
where $n_s$ is the superfluid density, and $\pi n_s e^2/(2m)$ is
the spectral weight under the $\delta$-functional piece of the
conductivity  in the superconducting state.

In practice, the integration up to an infinite frequency is hardly
possible, and more relevant issue for practical applications is
whether a sum rule is satisfied, at least approximately, for a
situation when there is a single electron band which  crosses
the Fermi level and is  well separated from other bands.
 Kubo considered this case in the same paper of 1957 and derived the
 expression   for the ``band'', or Kubo sum rule
\begin{equation}
 \int^{`\infty'}_{0} \, Re \,
\sigma (\Omega) \, d\Omega = W_K = \frac{\pi e^2}{2N}
\sum_{\vec{k}}\,\nabla_{\vec{k_x}}^2\varepsilon_{\vec{k}}\,n_{\vec{k}}
\label{eq:Kubo_sum}
\end{equation}
where $n_{\vec{k}}$ is the electronic distribution function and
$\varepsilon_{\vec{k}}$ is the band dispersion.
 Prime in the upper limit of the integration has the
practical implication that the upper limit is much larger than the
bandwidth of a given band which crosses the Fermi level, but
smaller than the frequencies of interband transitions.
Interactions with external objects, e.g., phonons or impurities,
and interactions between fermions are indirectly present in the
distribution function which is expressed via the full fermionic
Green's function as $n_{\vec{k}} = T \sum_m G({\vec{k}},
\omega_m)$.  For $\epsilon_k = k^2/2m$,
$\nabla_{\vec{k_x}}^2\varepsilon_{\vec{k}} = 1/m$, $W_K = \pi n
e^2/(2m)$, and Kubo sum rule reduces to Eq. (\ref{n_1}). In
general, however, $\varepsilon_{\vec{k}}$ is a lattice dispersion,
and Eqs. (\ref{n_1}) and (\ref{eq:Kubo_sum}) are different. Most
important, $W_K$ in Eq. (\ref{eq:Kubo_sum}) generally depends on
$T$ and on the state of the system because of $n_{\vec{k}}$.  In
this situation, the temperature evolution of the optical integral
does not reduce to a simple redistribution of the spectral weight
-- the whole spectral weight inside the conduction band changes
with $T$.  This issue was first studied in detail by Hirsch~
\cite{bib:jorge} who introduced the now-frequently-used notation
``violation of the conductivity sum rule''.

In reality, as already pointed out by Hirsch, there is no true
violation as the change of the total spectral weight in a given
band is compensated by an appropriate change of the spectral
weight in other bands such that the total spectral weight,
integrated over all bands, is conserved, as in Eq. (\ref{n_1}).
Still, non-conservation of the spectral weight within a given band
is an interesting phenomenon as the degree of non-conservation is
an indicator of relevant energy scales in the problem. Indeed,
when relevant energy scales are much smaller than the Fermi
energy, i.e., changes in the conductivity are confined to a near
vicinity of a Fermi surface (FS), one can expand $\varepsilon_k$
near $k_F$ as $\varepsilon_k = v_F (k-k_F) + (k-k_F)^2/(2m_B) +
O(k-k_F)^3$ and obtain $\nabla_{\vec{k_x}}^2\varepsilon_{\vec{k}}
\approx 1/m_B$ [this approximation is equivalent to approximating
the density of states (DOS) by a constant].  Then $W_K$ becomes $
\pi n e^2/(2m_B)$ which does not depend on temperature. The scale
of the temperature dependence of $W_K$ is then an indicator how
far in energy the changes in conductivity extend when, e.g., a
system evolves from a normal metal to a superconductor. Because
relevant energy scales increase with the interaction strength, the
temperature dependence of $W_K$ is also an indirect indicator of
whether a system is in a weak, intermediate, or strong coupling
regime.

In a conventional BCS superconductor the only relevant scales are
the superconducting gap $\Delta$ and the impurity scattering rate
$\Gamma$. Both  are generally much smaller than the Fermi energy,
so the optical integral should be almost $T$-independent, i.e.,
the spectral weight lost in a superconducting state at low
frequencies because of gap opening is completely recovered by the
zero-frequency $\delta$-function. In a clean limit, the weight
which goes into a $\delta-$function is recovered within
frequencies up to $4\Delta$. This is the essence of FGT sum rule
~\cite{bib:FGT1, bib:FGT2}. In a dirty limit, this scale is
larger, $O(\Gamma)$, but still $W_K$ is $T$-independent and there
was no ``violation of sum rule".

The issue of sum rule attracted substantial interest in the
studies of high $T_c$ cuprates~\cite{bib:BASOV, bib:MFL,
 bib:basov, bib:molegraaf, bib:optical int expt, bib:boris, bib:nicole, Carbone, Homes, Hwang,  Erik,
Ortolani, bib:Kin_kubo1, bib:KE relation, bib:opt int vio,
bib:cutoff07chu, bib:KE Hirsch, Toschi, Marsiglio, Benfatto} in
which pairing is without doubts a strong coupling phenomenon. From
a theoretical perspective, the interest  in this issue was
originally triggered by a similarity between $W_K$ and the kinetic
energy $K =2 \sum\,\varepsilon_{\vec{k}}
n_{\vec{k}}$.~\cite{bib:KE relation, bib:h2, bib:frank} For a model with a simple
tight binding cosine dispersion $\varepsilon_k \propto (\cos k_x +
\cos k_y)$, $\frac{d^2\,\varepsilon_{\vec{k}}}{d\,k_x^2}\thicksim
-\varepsilon_{\vec{k}}$ and $W_K = - K$. For a more complex
dispersion there is no exact relation between $W_K$ and $K$, but
several groups argued~ \cite{bib:Kin_kubo1, bib:Kin_kubo2,
bib:benfatto} that $W_K$ can still be regarded as a good monitor
for the changes in the kinetic energy. Now, in a BCS
superconductor, kinetic energy increases below $T_c$ because $n_k$
extends to higher frequencies (see Fig.\ref{fig:Dist fns}).
At strong coupling, $K$ not necessary increases  because of
opposite trend associated with the fermionic self-energy: fermions
are more mobile in the SCS due to less space for scattering at low
energies than they are in the NS. Model calculations show that
above some coupling strength, the kinetic energy decreases below
$T_c$~\cite{bib:sum rule frank}. While, as we said, there is no
one-to-one correspondence between $K$ and $W_K$, it is still
likely that, when $K$ decreases, $W_K$ increases.

A good amount of experimental effort has been put into addressing
the issue of the optical sum rule  in the
$c-$axis~\cite{bib:basov} and in-plane conductivities
~\cite{bib:molegraaf, bib:optical int expt, bib:boris, bib:nicole, Carbone, Homes, Hwang,  Erik, Ortolani} in
overdoped, optimally doped, and underdoped cuprates. The
experimental results demonstrated, above all, outstanding
achievements of experimental abilities as these groups managed to
detect the value of the optical integral with the accuracy of a
fraction of a percent. The analysis of the change of the optical
integral between normal and SCS is even more complex because one
has to (i) extend NS data to $T <T_c$ and (ii) measure superfluid
density with the same accuracy as the optical integral itself.

The analysis of  the optical integral showed that in overdoped
cuprates it definitely decreases below $T_c$, in consistency with
the expectations at weak coupling~\cite{bib:nicole}.  For
underdoped cuprates, all experimental groups agree that a relative
change of the optical integral below $T_c$ gets much smaller.
There is no agreement yet about the sign of the change of the
optical integral : Molegraaf \emph{et al.}\cite{bib:molegraaf} and
Santander-Syro \emph{et al.}\cite{bib:optical int expt}  argued
that the optical integral increases below $T_c$, while Boris
\emph{et al.}\cite{bib:boris} argued that it decreases.

Theoretical analysis of these results~\cite{bib:opt int vio,
bib:cutoff07chu, bib:norman pepin, Marsiglio, bib:benfatto} added
one more degree of complexity to the  issue.  It is tempting to
analyze the temperature dependence of $W_K$ and relate it to the
observed behavior of the optical integral, and some earlier
works\cite{bib:norman pepin, Marsiglio, bib:benfatto} followed
this route. In the experiments, however, optical conductivity is
integrated only up to a certain frequency $\w_c$, and the quantity
which is actually measured is \bea &&W(\w_c) = \int^{\w_c}_{0} \,
Re \,
\sigma (\Omega) \, d\Omega = W_K + f(\w_c) \nonumber \\
&&f(\w_c) = - \int_{\w_c}^{'\infty '} \, Re \, \sigma (\Omega) \,
d\Omega \label{n_4} \eea
The  Kubo formula, Eq.
(\ref{eq:Kubo_sum}) is obtained assuming that the second part
  is negligible. This is not guaranteed, however, as typical
 $\w_c \sim 1-2 eV$ are comparable to the bandwidth.

The differential sum rule $\Delta W$ is also a sum of two terms
\begin{equation}
\label{eq:vio_term} \Delta W(\omega_c) = \Delta W_K + \Delta
f(\w_c)
\end{equation}
where $\Delta W_K$ is the variation of the r.h.s. of Eq. \ref{eq:Kubo_sum}, and
 $\Delta f(\w_c)$ is the variation of the cutoff term.
 Because
conductivity changes with $T$ at all frequencies, $\Delta f(\w_c)$ also varies with temperature. It then becomes the issue
whether the experimentally observed  $\Delta W(\w_c)$ is predominantly due
to ``intrinsic''  $\Delta W_K$, or to $\Delta f(\w_c)$. [A third possibility is non-applicability of the
Kubo formula because of the close proximity of other bands, but we
will not dwell on this.]

\begin{figure}[htp]
\includegraphics*[width=.9\linewidth]{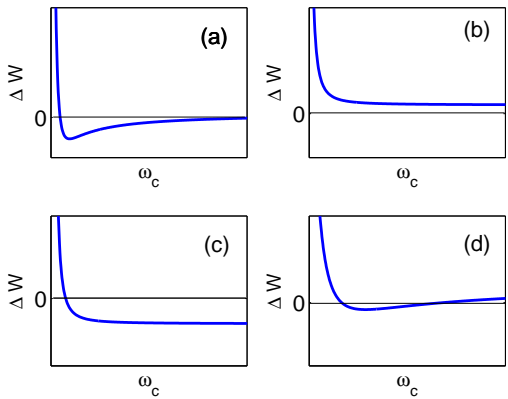}
\caption{\label{fig:schematic} Schematic behavior of
 $\Delta W$ vs $\omega_c$,  Eq. (\protect\ref{n_4}).
 The limiting value of $\Delta W$ at $\omega_c = \infty$ is
 $\Delta W_K$ given by Eq. (\protect\ref{eq:Kubo_sum})
Depending on the value of $\Delta W_K$, there
 can be either one sign change of $\Delta W$ (panels a and c),
or no sign changes (panel b), or two sign changes (panel d).}
\end{figure}

For the NS, previous works~\cite{bib:opt int vio, bib:cutoff07chu}
on particular models for the cuprates indicated that the origin of
the temperature dependence of $W(\w_c)$ is likely the $T$
dependence of the cutoff term $f(\omega_c)$.  Specifically, Norman
\emph{et. al.}\cite{bib:cutoff07chu} approximated a fermionic DOS
by a constant (in which case, as we said, $W_K$ does not depend on
temperature) and  analyzed the $T$ dependence of $W(\w_c)$ due to
the $T$ dependence of the cut-off term. They found a good
agreement with the experiments. This still does not solve the
problem fully as amount of the $T$ dependence  of $W_K$ in  the
same model but with a lattice dispersion has not been analyzed.
For a superconductor, which of the two terms contributes more,
remains an open issue.   At small frequencies, $\Delta
W(\omega_c)$ between a SCS and a NS is positive simply because
$\sigma (\Omega)$ in a SCS has a $\delta-$functional term. In the
models with a constant DOS, for which  $\Delta W_K =0$, previous
calculations~\cite{bib:opt int vio} show that $\Delta W(\w_c)$
changes sign at some $\w_c$, becomes negative at larger $\w_c$ and
approaches zero from a negative side. The frequency when  $\Delta
W(\w_c)$ changes sign is of order $\Delta$ at weak coupling,  but
increases as the coupling increases, and at large coupling becomes
comparable to a bandwidth ($\sim 1 eV$). At such frequencies the
approximation of a DOS by a constant is questionable at best, and
the behavior of $\Delta W(\w_c)$ should generally be influenced by
a nonzero $\Delta W_K$. In particular, the optical integral can
either remain positive for all frequencies below interband
transitions (for large enough positive $\Delta W_K$), or change
sign and remain negative  (for negative $\Delta W_K$). The first
behavior  would be consistent with Refs.
\onlinecite{bib:molegraaf, bib:optical int expt}, while the second
would be consistent with Ref. \onlinecite{bib:boris}.  $\Delta W$
can even show more exotic behavior with more than one sign change
(for a small positive $\Delta W_K$). We show various cases schematically in
 Fig.\ref{fig:schematic}.

In our work, we perform direct numerical calculations of optical
integrals at $T=0$ for a lattice dispersion extracted from ARPES of the
cuprates. The goal of our work is two-fold. First, we
perform calculations of the optical integral in the NS and analyze
how rapidly $W(\w_c)$ approaches $W_K$, in other words we check
how much of the Kubo sum is recovered up to the scale of the
bandwidth. Second, we analyze the difference between optical
integral in the SCS at $T=0$ and in the NS extrapolated to $T=0$
and compare the cut off effect $\Delta f(\omega_c)$ to $\Delta W_K$ term.
We also analyze the sign of $\Delta W(\w_c)$ at large
frequencies and discuss under what conditions theoretical
$W(\infty)$ increases in the SCS.

We perform calculations for four models. First is a conventional
BCS model with impurities (BCSI model). Second is an Einstein
boson (EB)  model of fermions interacting with a single Einstein
boson  whose propagator does not change between NS and SCS. These
two  cases will illustrate  a conventional idea of the spectral
weight in SCS being less than in NS. Then  we consider two more
sophisticated models: a phenomenological ``marginal Fermi liquid
with impurities'' (MFLI) model of Norman and
P\'{e}pin~\cite{bib:norman pepin}, and a microscopic collective
boson (CB) model~\cite{bib:fink} in which in the NS fermions
interact with a gapless continuum of bosonic excitations, but in a
$d-$wave SCS a gapless continuum splits into a resonance and a
gaped continuum. This model describes, in particular, interaction
of fermions with their own collective spin
fluctuations~\cite{bib:coupling to bos mode} via
\begin{equation}
\label{eq:NS_SCS_Self}\Sigma (k, \Omega)=
3 g^2 \int\,\frac{d\omega}{2\pi}\frac{d^2 q}{(2 \pi)^2} \, \chi(q, \omega)
G(k+q, \omega+\Omega)
\end{equation}
where $g$ is the spin-fermion coupling, and $\chi (q, \omega)$ is
the spin susceptibility whose dynamics changes between NS and SCS.

From our analysis we found that the introduction of a finite
fermionic bandwidth by means of a lattice has generally a
notable effect on both $W$ and $\Delta W$.
We found that for all models except for BCSI model, only
$70\%-80\%$ of the optical spectral weight is obtained by integrating up to the  bandwidth. In these three models, there also
 exists a wide range of $\w_c$ in which the behavior of $\Delta W(\w_c)$
 is due to variation of $\Delta
f(\w_c)$ which is dominant comparable to the $\Delta W_K$ term.
This dominance of  the cut off term is consistent with the
analysis in Refs. \onlinecite{bib:opt int vio, bib:cutoff07chu,
bib:sum rule mike_chu}.

We also found that for all models except for the original version
of the MFLI model the optical weight at the highest frequencies is
greater in the NS than in the SCS (i.e., $\Delta W <0$). This
observation is consistent with the findings of Abanov and
Chubukov~\cite{bib:coupling to bos mode},
 Benfatto \emph{et. al.}\cite{bib:benfatto}, and Karakozov and
Maksimov\cite{bib:karakozov}.
In the original version of the MFLI model \cite{bib:norman pepin}
 the spectral weight in SCS was found to be greater than in the NS
($\Delta W > 0$). We show that the behavior of $\Delta W (\w_c)$
in this model  crucially depends on how the fermionic self-energy
modeled to fit ARPES data in a NS is modified when a system
becomes a superconductor and can be of either sign. We also found,
however, that $\w_c$ at which $\Delta W$ becomes negative rapidly
increases with the coupling strength and at strong coupling
becomes comparable to the bandwidth. In the CB model, which, we
believe, is most appropriate for the application to the cuprates,
$\Delta W_K = \Delta W(\infty)$
 is quite small, and at strong coupling a negative $\Delta W (\w_c)$ up to
$\w_c \sim 1 eV$ is nearly compensated by the optical integral
between $\w_c$ and ``infinity'', which, in practice, is an energy
of interband transitions, which is roughly $2 eV$. This would be
consistent with Refs. \onlinecite{bib:molegraaf, bib:optical int
expt}.

We  begin with formulating our calculational basis in the next
section. Then we take up the four cases  and
consider in each case the extent to which the Kubo sum is satisfied
up to the order of bandwidth and the functional form and the sign of $\Delta W (\w_c)$.  The last section presents our conclusions.

%%%%%%%%%%%%%%%%%%%%%%%%%%%%%%%%%%%%%%%%%%%%%%%%%%%%%%%%%%%%%%%%%%%%%%
%%%%%%%%%%%%%%%%%%%%%%%%%%%%%%%%%%%%%%%%%%%%%%%%%%%%%%%%%%%%%%%%%%%%%%
\section{Optical Integral in Normal and Superconducting states}

\begin{figure}[htp]
\includegraphics*[width=.9\linewidth]{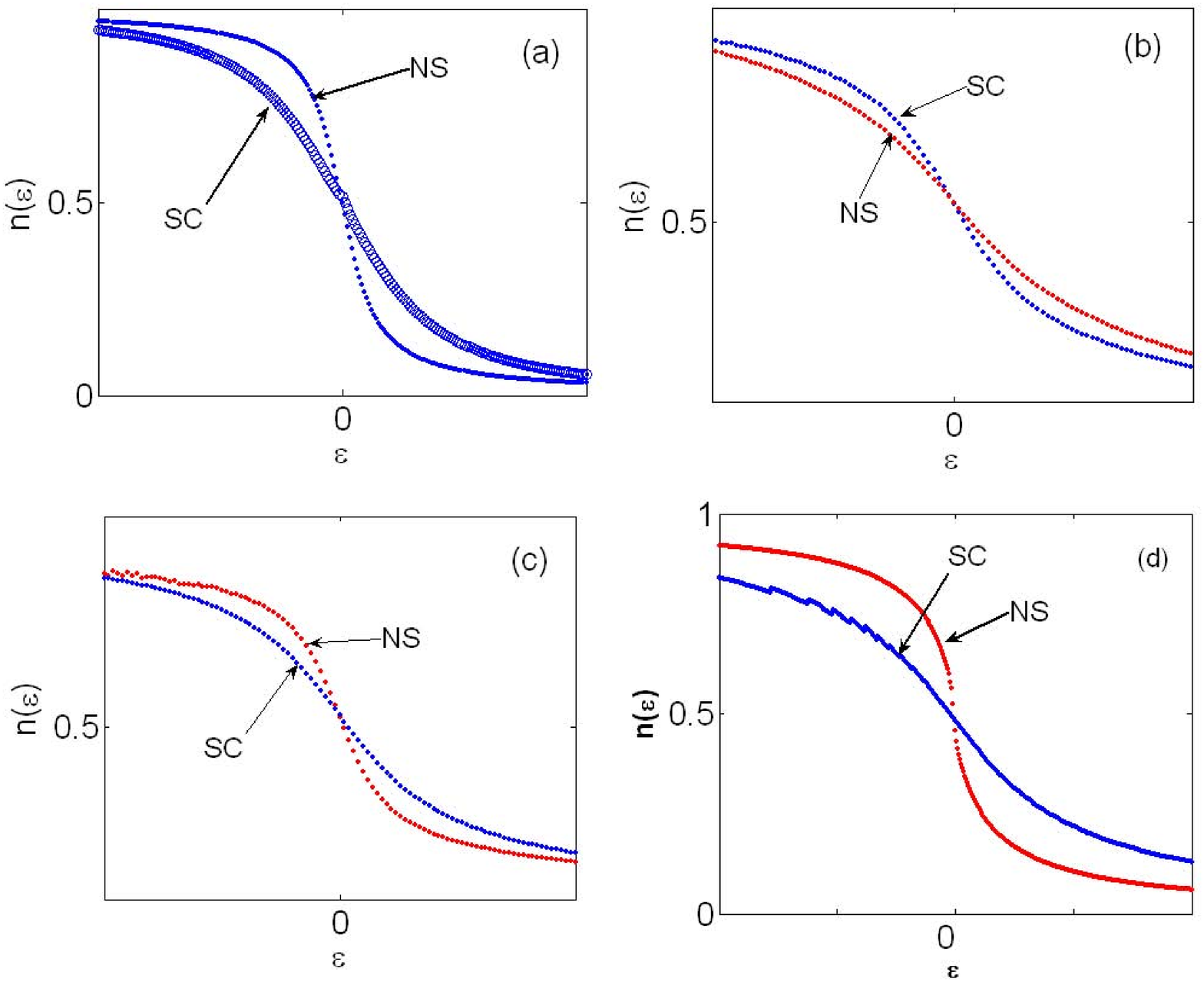}
\caption{\label{fig:Dist fns}Distribution functions in four cases
(a) BCSI model, where one can see that for $\varepsilon>0$,
SC$>$NS implying KE increases in the SCS. (b) The original MFLI
model of Ref. \onlinecite{bib:norman pepin}, where for
$\varepsilon>0$, SC$<$NS, implying KE decreases in the SCS. (c) Our
version of MFLI model (see text) and
 (d) the CB model.  In both cases, SC$>$NS, implying KE
increases in the SCS. Observe that in  the impurity-free CB model
there is no jump in $n (\epsilon)$ indicating lack of fermionic coherence. This is consistent with ARPES\cite{bib:no quasi NS}}
\end{figure}

The generic formalism of the computation of the optical
conductivity and the optical integral has been discussed several
times in the literature~\cite{bib:opt int vio, bib:cutoff07chu,
bib:sum rule frank, bib:KE Hirsch, Benfatto} and we just list the
formulas that we used in our computations. The conductivity
$\sigma (\Omega)$ and the optical integral $W(\w_c)$
 are given by (see for example Ref.
\onlinecite{bib:cond_def}).
\begin{subequations}
\begin{align}
\label{eq:re_cond} \sigma'(\Omega) & =
Im\left[-\frac{\Pi(\Omega)}{\Omega+i\delta}\right]
= -\frac{\Pi''(\Omega)}{\Omega}\;+\;\pi\delta(\Omega)\,\Pi'(\Omega)\\
\label{eq:opt_int}W(\omega_c) & = \int^{\omega_c}_{0} \,
\sigma'(\Omega) \, d\Omega = -\int^{\omega_c}_{0+} \,
\frac{\Pi''(\Omega)}{\Omega} \, d\Omega \;+\; \frac{\pi}{2}\Pi'(0)
\end{align}
\end{subequations}
where `$X'$' and `$X''$' stand for real and imaginary parts of
$X$. We will restrict with $T=0$. The polarization operator
$\Pi(\Omega)$ is (see Ref. \onlinecite{bib:pi_def})
\begin{widetext}
\begin{subequations}
\begin{align}
\label{eq:pi_mat}\Pi(i\Omega) &=T\sum_\omega\,\sum_{\vec{k}}\,
(\nabla_{\vec{k}}\varepsilon_{\vec{k}})^2\, \left(
    G(i\omega,\vec{k})G(i\omega+i\Omega,\vec{k})\,+\,
    F(i\omega,\vec{k})F(i\omega+i\Omega,\vec{k})
\right)\\
\label{eq:pi_im}\Pi''(\Omega)&=-\frac{1}{\pi}
\sum_{\vec{k}}\,(\nabla_{\vec{k}}\varepsilon_{\vec{k}})^2
\int^0_{-\Omega}\, d\omega \, \left(
    G''(\omega,\vec{k})G''(\omega+\Omega,\vec{k})\,+\,
    F''(\omega,\vec{k})F''(\omega+\Omega,\vec{k})
\right)\\
\label{eq:pi_re}\Pi'(\Omega)&= \frac{1}{\pi^2}
\sum_{\vec{k}}\,(\nabla_{\vec{k}}\varepsilon_{\vec{k}})^2
\int'\int'\,dx\,dy\, \left(
    G''(x,\vec{k})G''(y,\vec{k})\,+\,
    F''(x,\vec{k})F''(y,\vec{k})
\right)\, \frac{
    n_F(y)-n_F(x)}{y-x}
\end{align}
\end{subequations}
\end{widetext}

where $\int'$ denotes the principal value of the integral,
$\sum_{\vec{k}}$ is understood to be
$\frac{1}{N}\sum_{\vec{k}}$,($N$ is the number of lattice sites),
 $n_F(x)$ is the Fermi function which is a step function at zero
temperature, $G$ and $F$ are the normal and anomalous Greens functions.
 given by
\cite{bib:greens functions}
\begin{subequations}
\begin{align}
\label{eq:NS_G}\text{For a NS,\    }&
G(\omega,\vec{k})=\frac{1}{\omega-\Sigma(k, \omega)-\varepsilon_{\vec{k}}+i\delta}\\
\label{eq:SCS_G}\text{For a SCS,\  }&
G(\omega,\vec{k})=\frac{Z_{k,\omega} \omega+\varepsilon_{\vec{k}}}
{Z^2_{k,\omega} (\omega^2-\Delta^2_{k,\omega})-\varepsilon_{\vec{k}}^2+i\delta sgn(\omega)}\\
\label{eq:SCS_F}&F(\omega,\vec{k})=\frac{Z_{k,\omega} \Delta_{k,\omega}}
{Z^2_{k,\omega} (\omega^2-\Delta^2_{k,\omega})-\varepsilon_{\vec{k}}^2+i\delta
sgn(\omega)}
\end{align}
\end{subequations}
where $Z_{k,\omega} =1-\frac{\Sigma (k,\omega)}{\omega}$, and
$\Delta_{k,\omega}$, is the SC gap. Following earlier
works~\cite{bib:sum rule mike_chu, bib:fink}, we  assume that the
fermionic self-energy $\Sigma (k, \omega)$ predominantly depends
on frequency and approximate $\Sigma (k, \omega) \approx \Sigma
(\omega)$ and also neglect the frequency dependence of the gap,
i.e., approximate  $\Delta_{k,\omega}$ by a $d-$wave $\Delta_k$.
The lattice dispersion $\varepsilon_{\vec{k}}$  is taken from Ref.
\onlinecite{bib:dispersion}. To calculate $W_K$, one has to
evaluate the Kubo term in Eq.\ref{eq:Kubo_sum} wherein  the
distribution function $n_{\vec{k}}$, is calculated from
\begin{equation}
\label{eq:n_k} n(\varepsilon_{\vec{k}})=-2\int_{-\infty}^0
\frac{d\omega}{2\pi} \, G''(\omega,\vec{k})
\end{equation}
The $2$ is due to the trace over spin indices. We show  the
distribution functions in the NS and SCS under different
circumstances in Fig \ref{fig:Dist fns}.

The $\vec{k}$-summation is done over first Brillouin zone for a
2-D lattice with a 62x62 grid. The frequency integrals are done
analytically wherever possible, otherwise performed using
Simpson's rule for all regular parts. Contributions from the poles
are computed separately  using Cauchy's theorem. For comparison,
in all four cases we also calculated FGT sum rule by replacing
$\int d^2 k = d \Omega_k d \epsilon_k \nu_{\epsilon_k,\Omega_k}$
and keeping $\nu$ constant.   We remind that the
 FGT is the result when one assumes that the integral in $W(\w_c)$
predominantly comes  from  a narrow region around the Fermi
surface.

We will first use Eq
\ref{eq:Kubo_sum} and compute $W_K$ in NS and SCS. This will tell us
 about the magnitude of $\Delta W (\w_c = \infty)$.  We
next compute the conductivity $\sigma(\omega)$ using the equations
listed above, find $W(\w_c)$ and $\Delta W (\omega_c)$ and compare
$\Delta f(\w_c)$ and $\Delta W_K$.

For simplicity and also for comparisons with earlier studies, for
BCSI, EB, and MFLI models we assumed that the gap is just a
constant along the FS. For CB model, we used a $d-$wave gap and
included into consideration the fact that, if a CB is a spin
fluctuation, its propagator develops a resonance when the pairing
gap is $d-$wave.

%%%%%%%%%%%%%%%%%%%%%%%%%%%%%%%%%%%%%%%%%%%%%%%%%%%%%%%%%%%%%%%%%%%%%%
\subsection{The BCS case}

In BCS theory the quantity $Z(\omega)$ is
given by
\begin{equation}
\label{eq:BCS_Z}
Z_{BCSI} (\omega)=1+\frac{\Gamma}{\sqrt{\Delta^2-(\omega+i\delta)^2}}
\end{equation}
and
\beq
\Sigma_{BCSI} (\omega) = \omega \left(Z(\omega)-1\right) = i \Gamma \frac{\omega}{\sqrt{(\omega+i\delta)^2 - \Delta^2}}
\label{eq:sigma_bcs}
\eeq
This is consistent with having in the NS, $\Sigma=i\Gamma$ in
accordance with Eq \ref{eq:NS_SCS_Self}. In the SCS, $\Sigma (\omega)$ is
purely imaginary for $\omega > \Delta$ and purely real for $\omega < \Delta$.
The self-energy has a square-root singularity at $\omega = \Delta$.

It is worth noting that Eq.\ref{eq:sigma_bcs} is derived from the
integration over infinite band. If one uses
Eq.\ref{eq:NS_SCS_Self} for finite band, Eq.\ref{eq:sigma_bcs}
acquires an additional frequency dependence at large frequencies
of the order of bandwidth (the low frequency structure still
remains the same as in Eq.\ref{eq:sigma_bcs}). In principle, in a
fully self-consistent analysis, one should indeed evaluate the
self-energy using a finite bandwidth. In practice, however, the
self-energy at frequencies of order bandwidth is generally much
smaller than $\omega$ and contribute very little to optical
conductivity which predominantly comes from frequencies where the
self-energy is comparable or even larger than $\omega$. Keeping
this in mind, below we will continue with the form of self-energy
derived form infinite band. We use the same argument for all four
models for the  self-energy.

For completeness, we first present some well known results
about the conductivity and optical integral for a constant DOS and then
extend the
discussion to the case where the same calculations are done in the
presence of a particular lattice dispersion.

\begin{figure}[htp]
\includegraphics*[width=.7\linewidth]{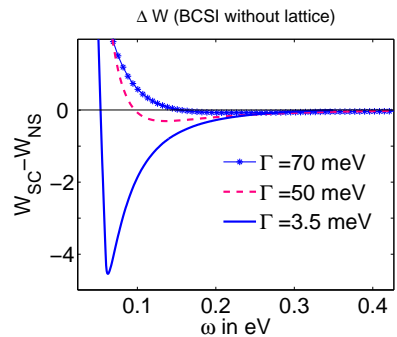}
\caption{\label{fig:BCS_OpDf_nolat} The BCSI case with a dispersion linearized  around the Fermi surface.
Evolution of the difference of
optical integrals in the SCS and the NS with the upper cut-off $\omega_c$
 Observe that the zero crossing
point increases with impurity scattering rate $\Gamma$ and also the `dip' spreads out with
increasing $\Gamma$. $\Delta=30\,meV$}
\end{figure}

For a constant DOS, $\Delta W(\omega_c)=W_{SC}(\omega_c)-W_{NS}(\omega_c)$
 is zero at
$\omega_c=\infty$ and Kubo sum rule reduces to FGT sum rule.
In Fig. \ref{fig:BCS_OpDf_nolat} we plot for this case $\Delta W(\omega_c)$ as
a function of the cutoff $\omega_c$ for different $\Gamma 's$.
The plot shows the two well known
features: zero-crossing point is
 below $2\Delta$ in the clean limit $\Gamma << \Delta$ and is
roughly $2\Gamma$ in the dirty limit \cite{bib:artem_BCSI,bib:opt
int vio} The magnitude of the  `dip'  decreases quite rapidly with
increasing $\Gamma$. Still, there is  always a point of zero
crossing and $\Delta W (\w_c)$ at large $\w_c$ approaches zero
from below.

\begin{figure}[htp]
\includegraphics*[width=.7\linewidth]{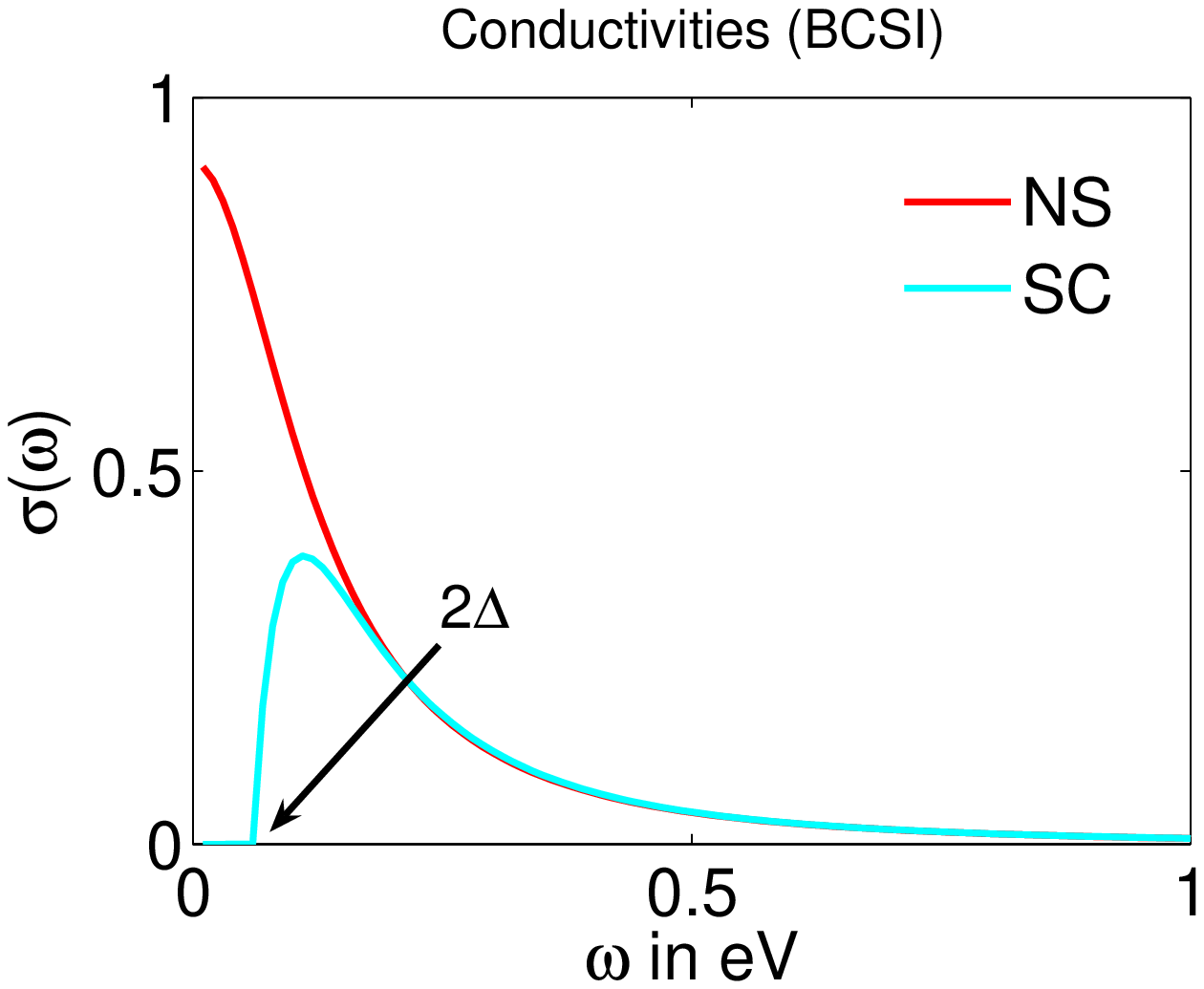}
\hfill
\includegraphics*[width=.7\linewidth]{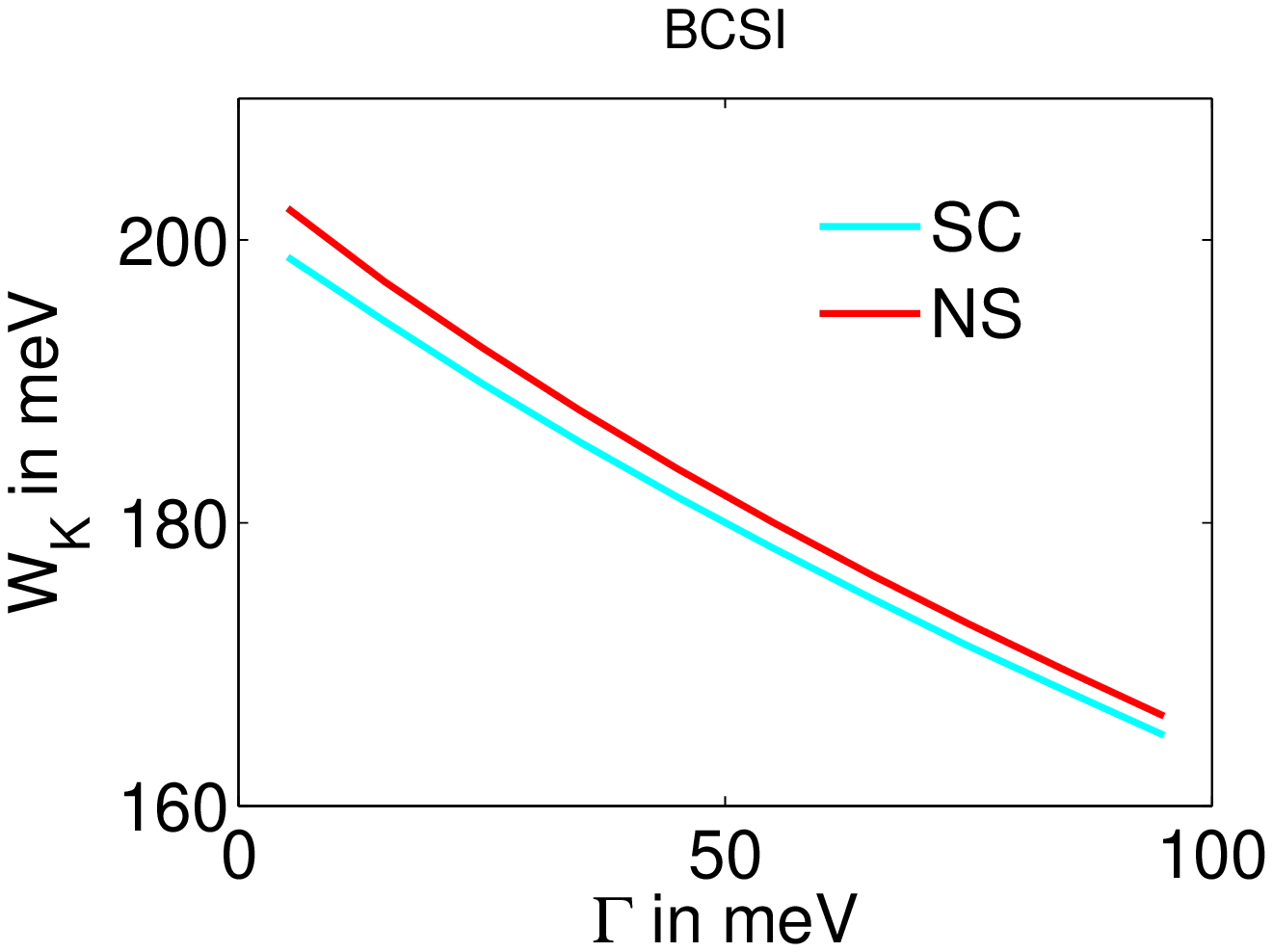}
\caption{\label{fig:BCS_KS_Cond}Top - a  conductivity plot
for the BCSI case in the presence of a lattice. The
parameters are $\Delta=30\,meV$, $\Gamma=3.5\,meV$. Bottom -- the
behavior of Kubo sums. Note that (a) the
spectral weight in the NS is always greater in the SCS, (b) the spectral weight decreases with $\Gamma$, and  (c) the difference between  NS
and SCS decreases as $\Gamma$ increases.}
\end{figure}

\begin{figure}[htp]
\includegraphics*[width=.7\linewidth]{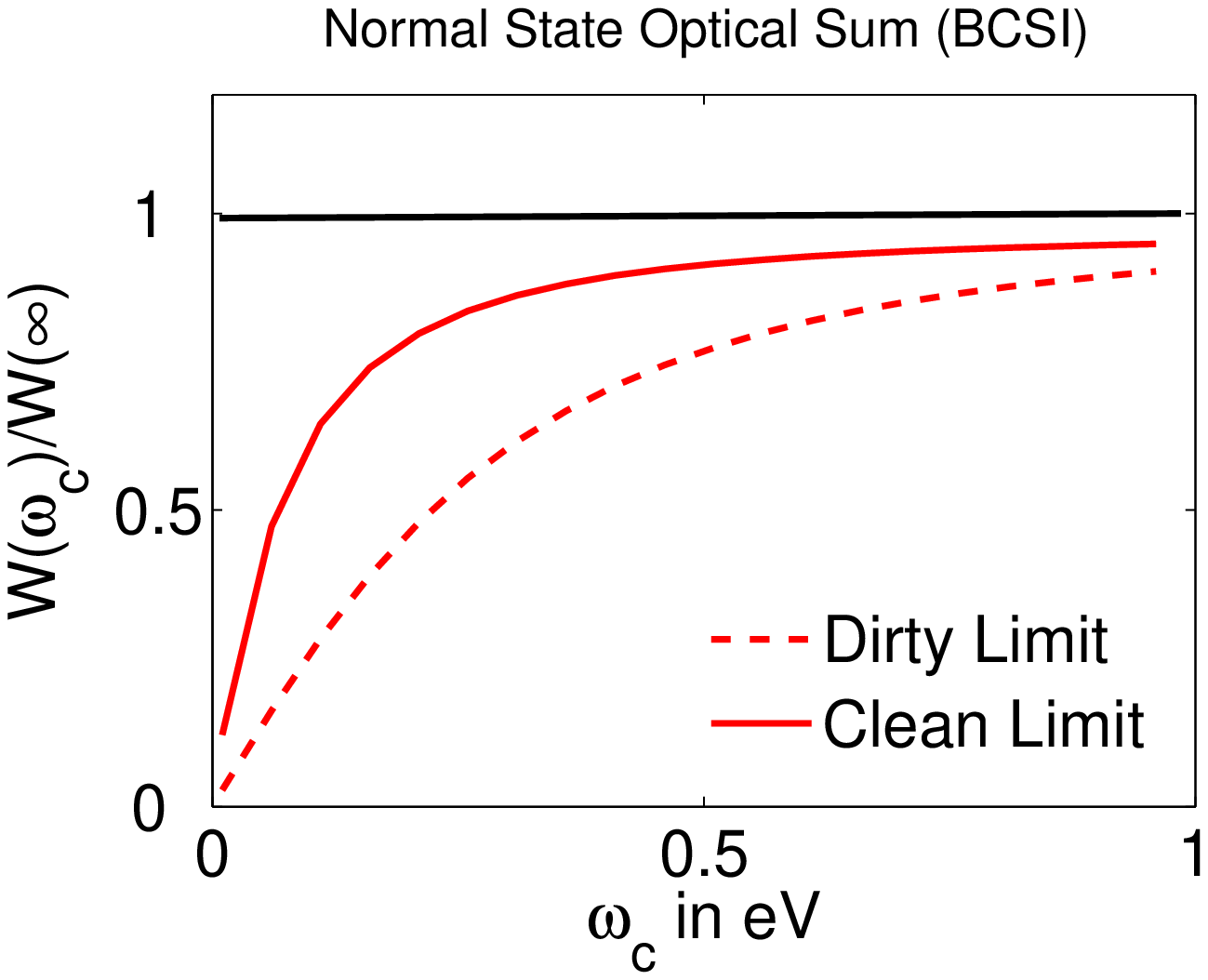}
\hfill
\includegraphics*[width=.7\linewidth]{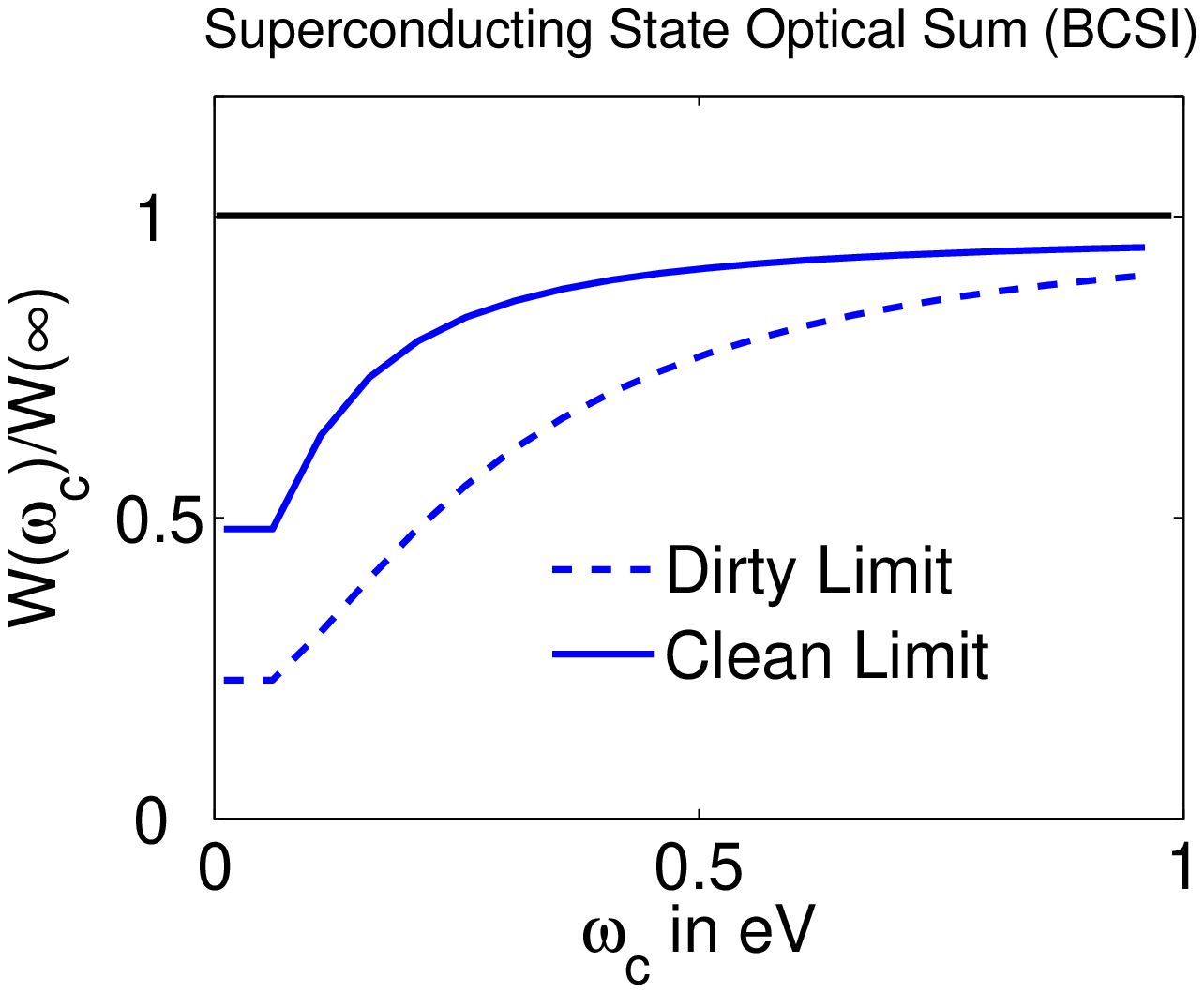}
\caption{\label{fig:BCS_OS}The evolution of optical integral in
NS(top) and SCS(bottom) for BCSI case. Plots are made
for clean limit (solid lines, $\Gamma=3.5\,meV$) and dirty limit
(dashed lines, $\Gamma=150\,meV$) for $\Delta=30\,meV$. Observe that
(a) $W(0)=0$ in the NS, but has a non-zero value in the SCS
 because of the $\delta$-function (this value decreases in the dirty limit),
 and (b) the flat region in the SCS is
due to the fact that $\sigma'(\w)=0$ for $\Omega<2\Delta$. Also
note that $\sim 90-95\%$ of the spectral weight is recovered up to $1 eV$}
\end{figure}

\begin{figure}[htp]
\includegraphics*[width=.7\linewidth]{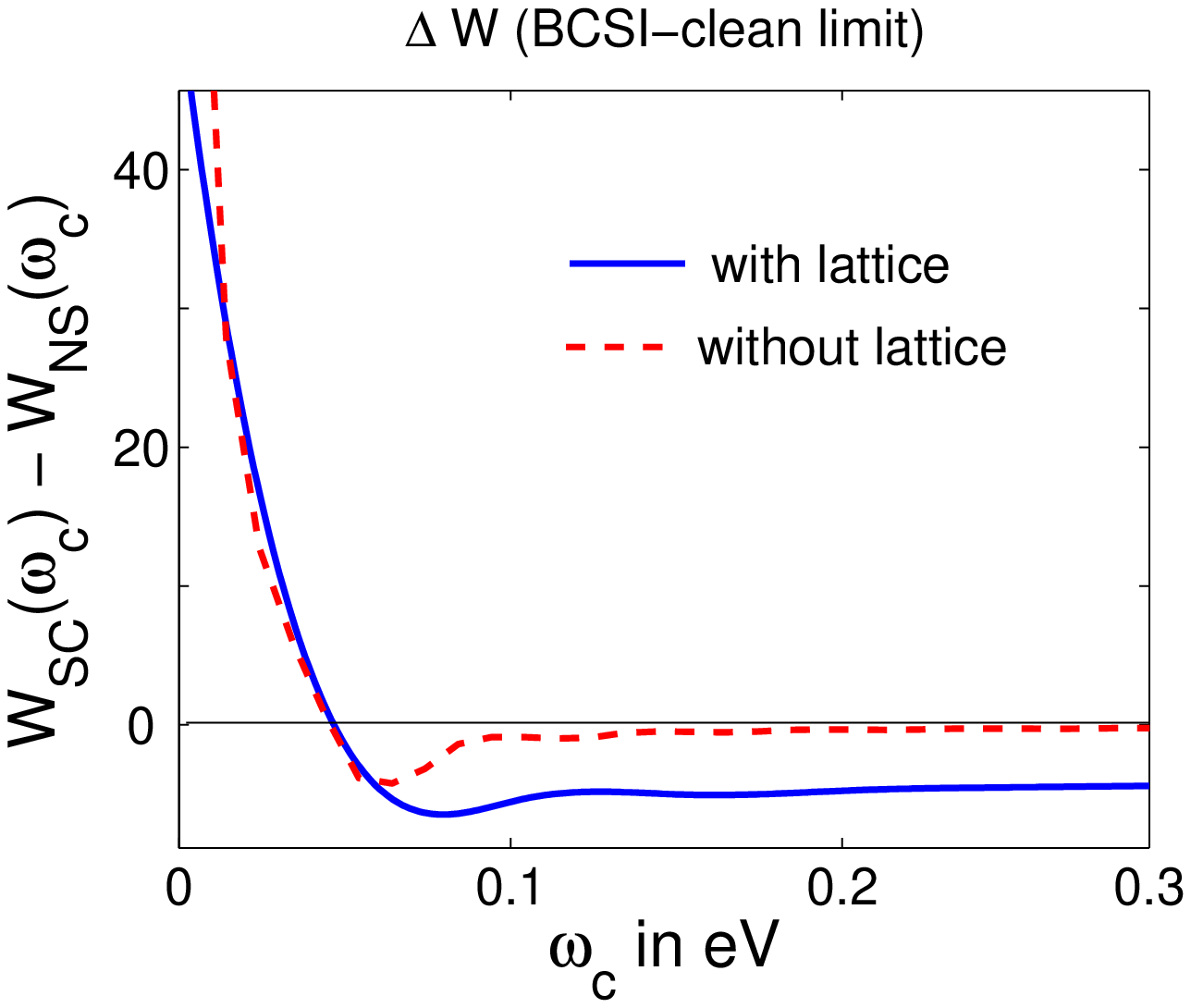}
\hfill
\includegraphics*[width=.7\linewidth]{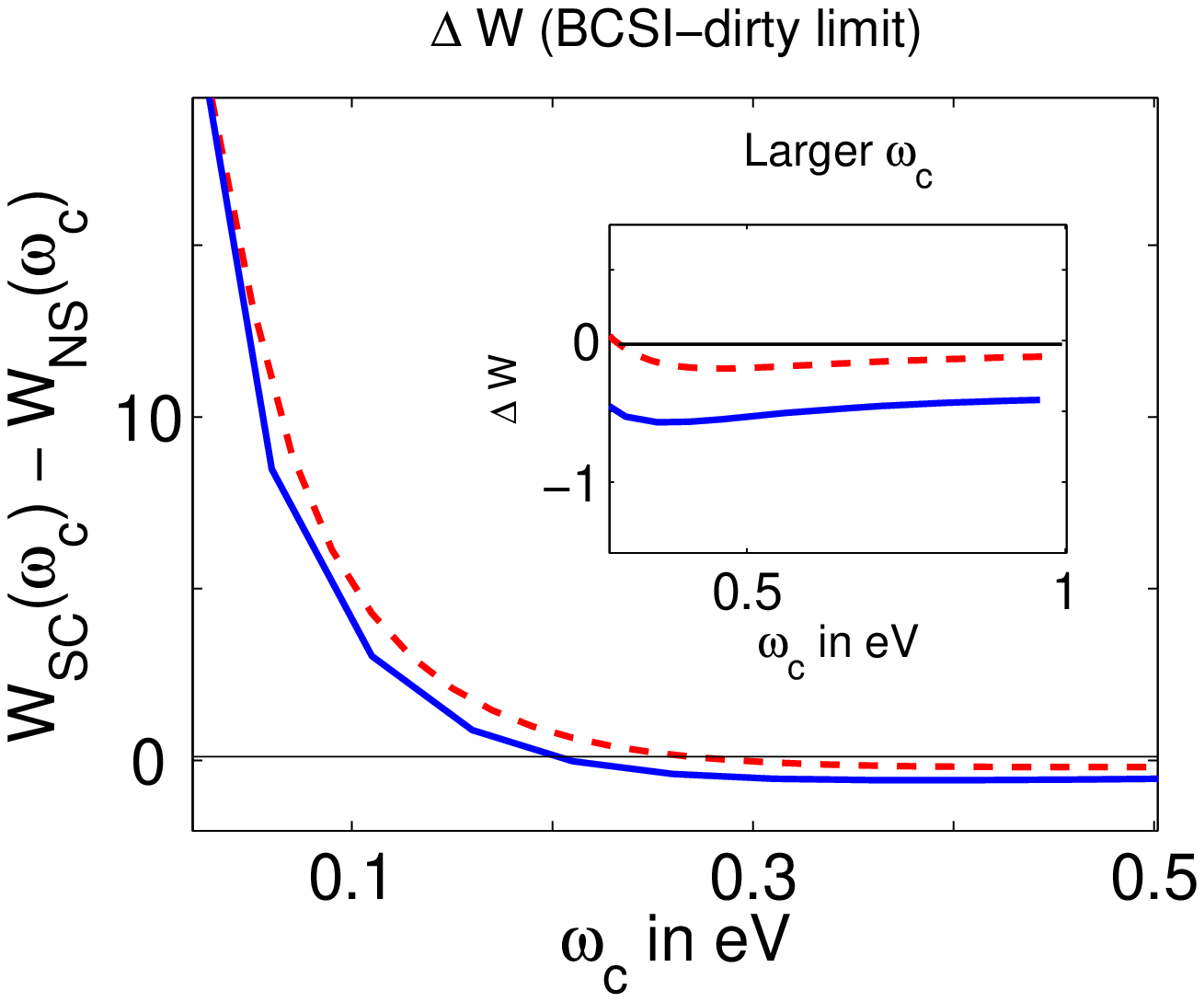}
\caption{\label{fig:BCS_optdiff_comp}Evolution of $\Delta W$ in
the presence of a lattice (solid line) compared with the case of
no lattice(a constant DOS, dashed line)  for clean and dirty
limits. $\Delta = 30\,meV$, $\Gamma=3.5 \,meV$ (clean limit),
$\Gamma=150 \,meV$ (dirty limit) }
\end{figure}

We now perform the same calculations in the presence of lattice
dispersion. The results are summarized in Figs
\ref{fig:BCS_KS_Cond},\ref{fig:BCS_OS}, and
\ref{fig:BCS_optdiff_comp}.

Fig \ref{fig:BCS_KS_Cond} shows conductivities $\sigma (\omega)$
in the NS and the SCS and
Kubo sums $W_K$ plotted against impurity scattering $\Gamma$.
We see that the
optical integral  in the NS is always greater than in the SCS. The
 negative sign of $\Delta W_K$  is simply the consequence of the
 fact that  $n_k$ is larger in the NS for $\epsilon_k <0$ and smaller for
$\epsilon_k <0$, and
$\nabla^2\varepsilon_{\vec{k}}$  closely follows
$-\varepsilon_{\vec{k}}$  for our
choice of dispersion \cite{bib:dispersion}),
Hence  $n_k$ is larger in the NS for $\nabla^2\varepsilon_{\vec{k}} >0$ and smaller for $\nabla^2\varepsilon_{\vec{k}} <0$ and  the Kubo sum rule, which is the integral of the product of $n_k$ and $\nabla^2\varepsilon_{\vec{k}}$ (Eq. \ref{eq:Kubo_sum}), is larger in the normal state.

We also see from Fig. \ref{fig:BCS_KS_Cond} that  $\Delta W_K$
decreases with $\Gamma$  reflecting the fact that with too much
impurity scattering there is little difference in $n_k$ between NS
and SCS.

Fig \ref{fig:BCS_OS} shows the optical sum in NS and SCS in clean
and dirty limits (the parameters are stated in the figure). This
plot shows that the Kubo sums are almost completely recovered
by integrating up to the bandwidth of $1 eV$: the recovery is
 $95\%$ in the clean limit and $\thicksim 90\%$ in the dirty limit.
In Fig \ref{fig:BCS_optdiff_comp} we plot $\Delta W (\w_c)$ as a
function of $\w_c$
 in clean and dirty limits. $\Delta W (\infty)$ is now non-zero, in agreement
 with Fig. \ref{fig:BCS_KS_Cond} and we also see that there is little variation
 of $\Delta W (\omega_c)$ at above $0.1-0.3 eV$ what implies that for
 larger $\w_c$, $\Delta W (\w_c) \approx \Delta W_K >> \Delta f(\w_c)$.

To make this more quantitative, we compare in
 Fig. \ref{fig:BCS_optdiff_comp} $\Delta W (\w_c)$ obtained
 for a constant DOS, when $\Delta W (\w_c) = \Delta f (\w_c)$,
 and for the actual lattice dispersion, when $\Delta W (\w_c) = \Delta W_K +
 \Delta f (\w_c)$. In the clean limit there is obviously little cutoff
 dependence beyond $0.1 eV$, i.e., $\Delta f (\w_c)$ is truly small,
 and the difference between the two cases is just $\Delta W_K$.
In the dirty limit, the situation is similar, but there is obviously
 more variation with $\omega_c$, and $\Delta f (\w_c)$ becomes truly small
 only above $0.3 eV$. Note also that the position of the dip in
 $\Delta W (\w_c)$ in the clean limit is at a larger $\w_c$ in the presence of
 the lattice than in a continuum.

%%%%%%%%%%%%%%%%%%%%%%%%%%%%%%%%%%%%%%%%%%%%%%%%%%%%%%%%%%%%%%%%%%%%%%
\subsection{The Einstein boson model}

We next consider the case
of electrons interacting with a single boson mode which by itself is not affected by superconductivity. The primary candidate for such mode is an optical phonon.
 The imaginary part
of the NS self energy has been discussed numerous times in the literature.
We make one simplifying assumption -- approximate the
 DOS by a constant in calculating fermionic self-energy.
 We will, however, keep the full lattice dispersion in the calculations of the
 optical integral. The advantage of this approximation is that the self-energy
 can be computed analytically. The full self-energy obtained with the lattice
 dispersion is more involved and can only be obtained numerically, but its
 structure is quite similar to the one obtained with a constant DOS.

The self-energy for a constant DOS is given by
\begin{equation}
\Sigma (i\omega) = -\frac{i}{2\pi} \lambda_n \int d \epsilon_k d
(i\Omega) \chi (i\Omega) G (\epsilon_k, i\omega + i\Omega)
\end{equation}
where
\begin{equation}
\chi (i\Omega) = \frac{\omega^2_0}{\omega^2_0 - (i\Omega)^2}
\end{equation}
and  $\lambda_n$ is a dimensionless electron-boson  coupling.
Integrating and transforming to real frequencies, we obtain
\[
\label{eq:PM_ImSE_NS}
\Sigma''(\omega)=-\frac{\pi}{2}\,\lambda_n\omega_o\,\Theta(|\omega|-\omega_o)
\]
\begin{equation}
\label{eq:PM_ReSE_NS}
\Sigma'(\omega)=-\frac{1}{2}\,\lambda_n\omega_o\,log\left|
\frac{\omega+\omega_o}{\omega-\omega_o} \right|
\end{equation}
 In the SCS, we obtain for $\omega<0$
\[
\label{eq:PM_ImSE_SC}
\Sigma''(\omega)=-\frac{\pi}{2}\,\lambda_n\omega_o\,Re \left(
\frac{\omega+\omega_o}{\sqrt{(\omega+\omega_o)^2-\Delta^2}}\right)
\]
\begin{equation}
\label{eq:PM_ReSE_SC}
\Sigma'(\omega)=-\frac{1}{2}\,\lambda_n\omega_o\,Re \int\,
d\omega'
\frac{1}{\omega_o^2-\omega'^2-i\delta}\frac{\omega+\omega'}{\sqrt{(\omega+\omega')^2-\Delta^2}}
\end{equation}
Observe that  $\Sigma''(\omega)$ is no-zero only for $\omega\, <\,
- \omega_o-\Delta$. Also, although it does not straightforwardly
follow from Eq. \ref{eq:PM_ReSE_SC}, but real and imaginary parts
of the self-energy do satisfy
 $\Sigma'(\omega)=-\Sigma'(-\omega)$ and
$\Sigma''(\omega)=\Sigma''(-\omega)$.

\begin{figure}[htp]
\includegraphics*[width=.7\linewidth]{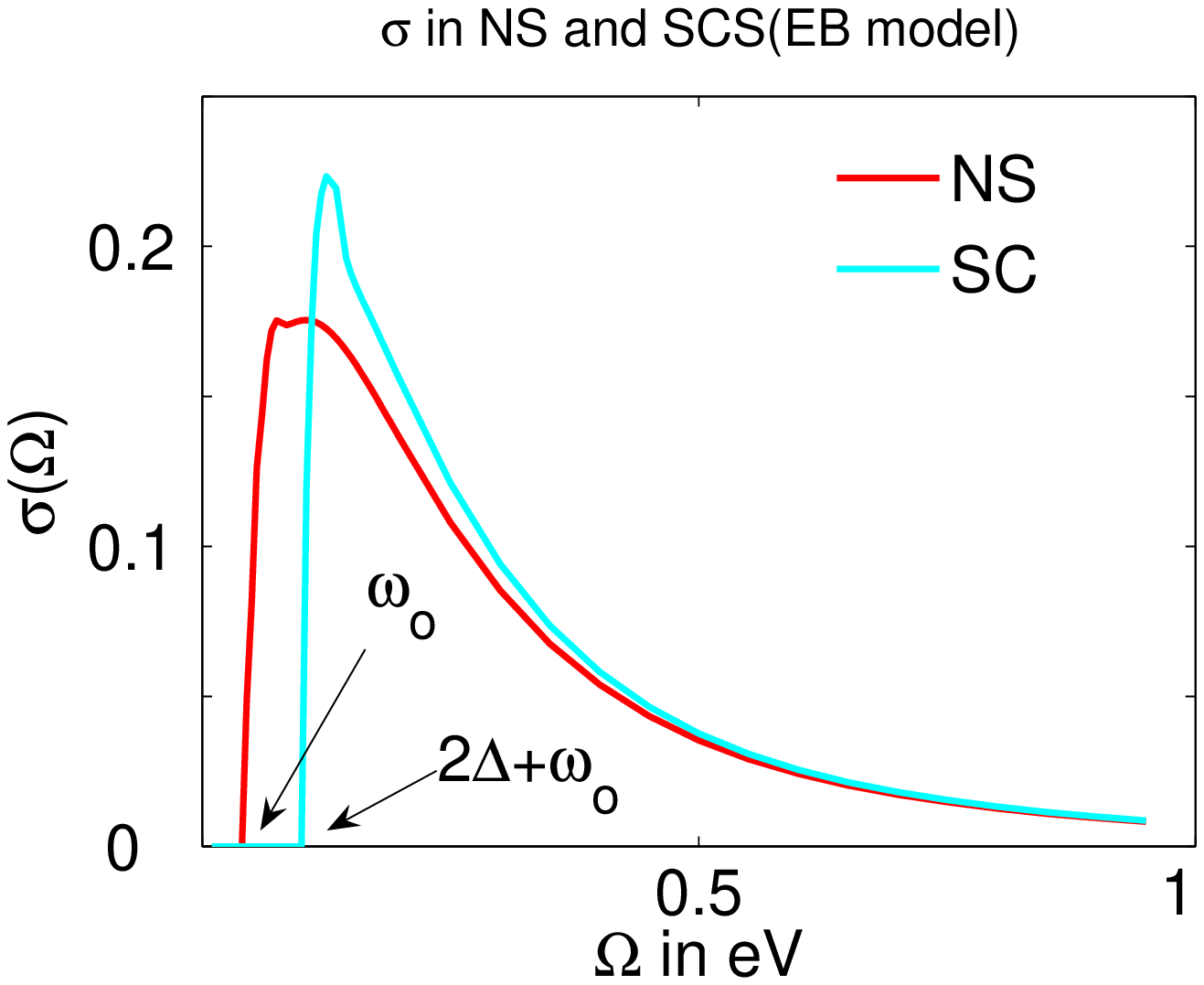}
\hfill
\includegraphics*[width=.7\linewidth]{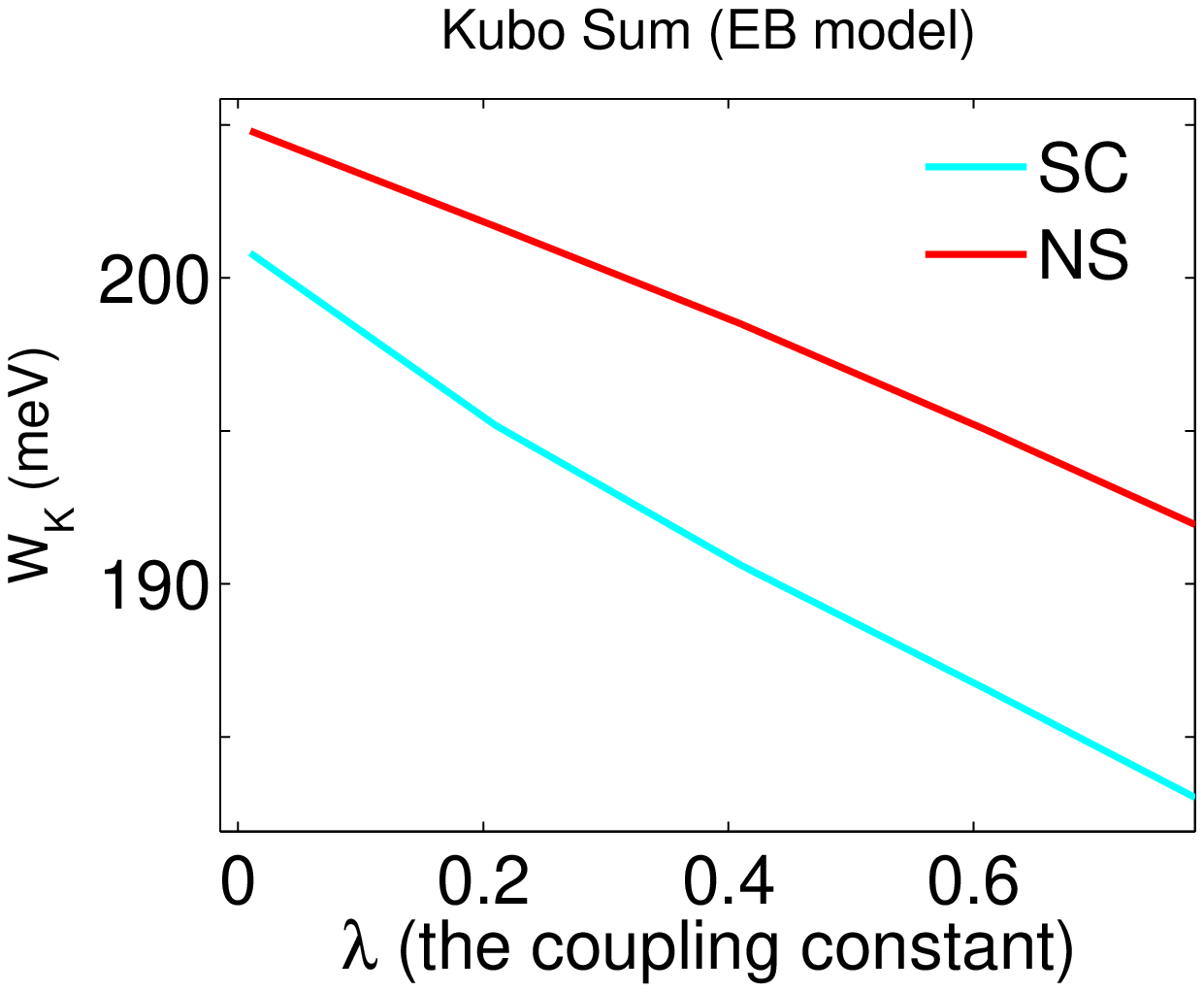}
\caption{\label{fig:PM_KS_Cond}Top- conductivities in the NS and the SCS
for the EB model. The conductivity in the NS vanishes below $\omega_0$ because of no phase space for scattering. Bottom - Kubo sums as a function
of coupling. Observe that $W_K$ in the SCS is below that in the NS. We set $\omega_o = 40\,meV$, $\Delta=30\,meV$, $\lambda=.5$}
\end{figure}

\begin{figure}[htp]
\includegraphics*[width=.7\linewidth]{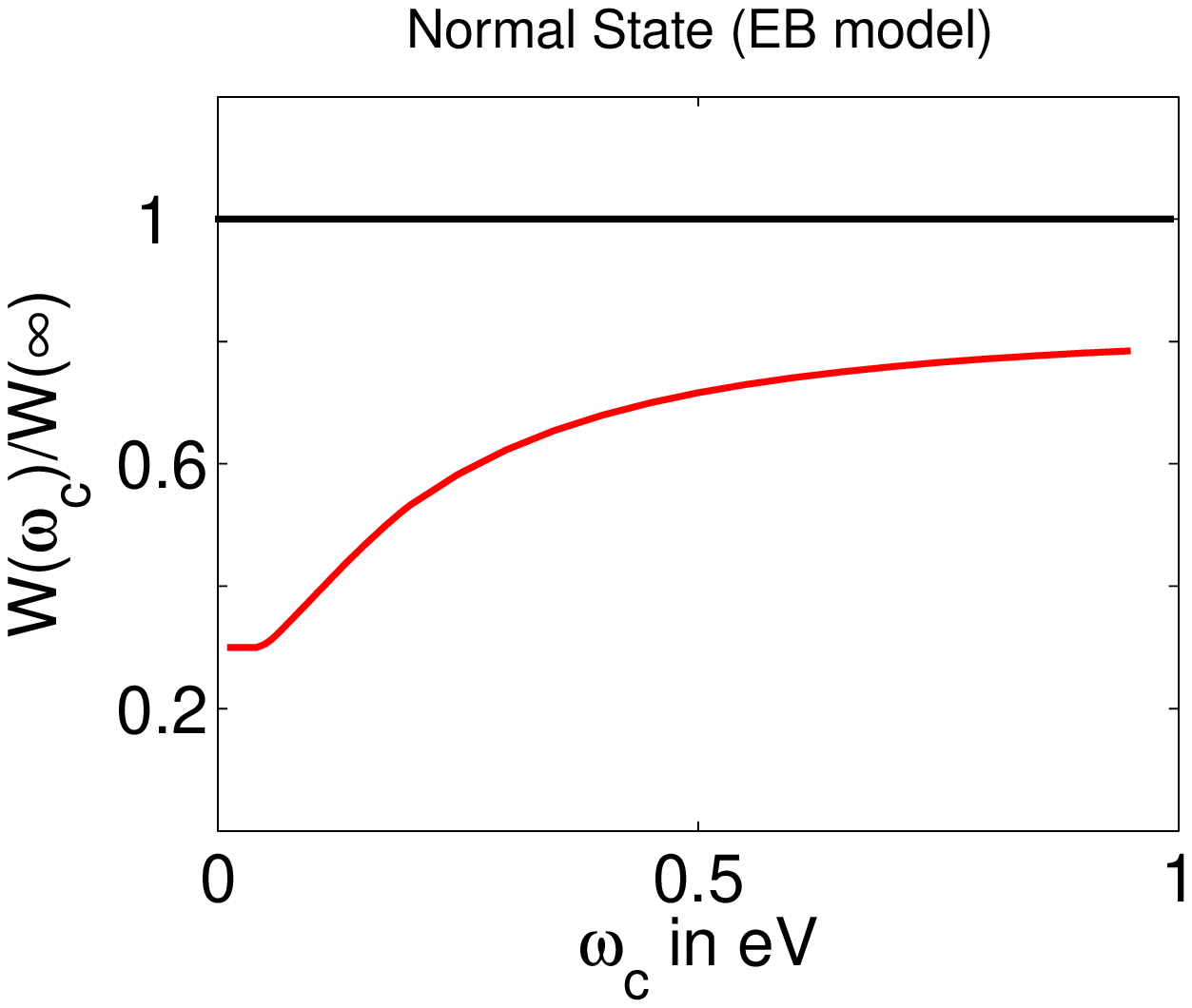}
\hfill
\includegraphics*[width=.7\linewidth]{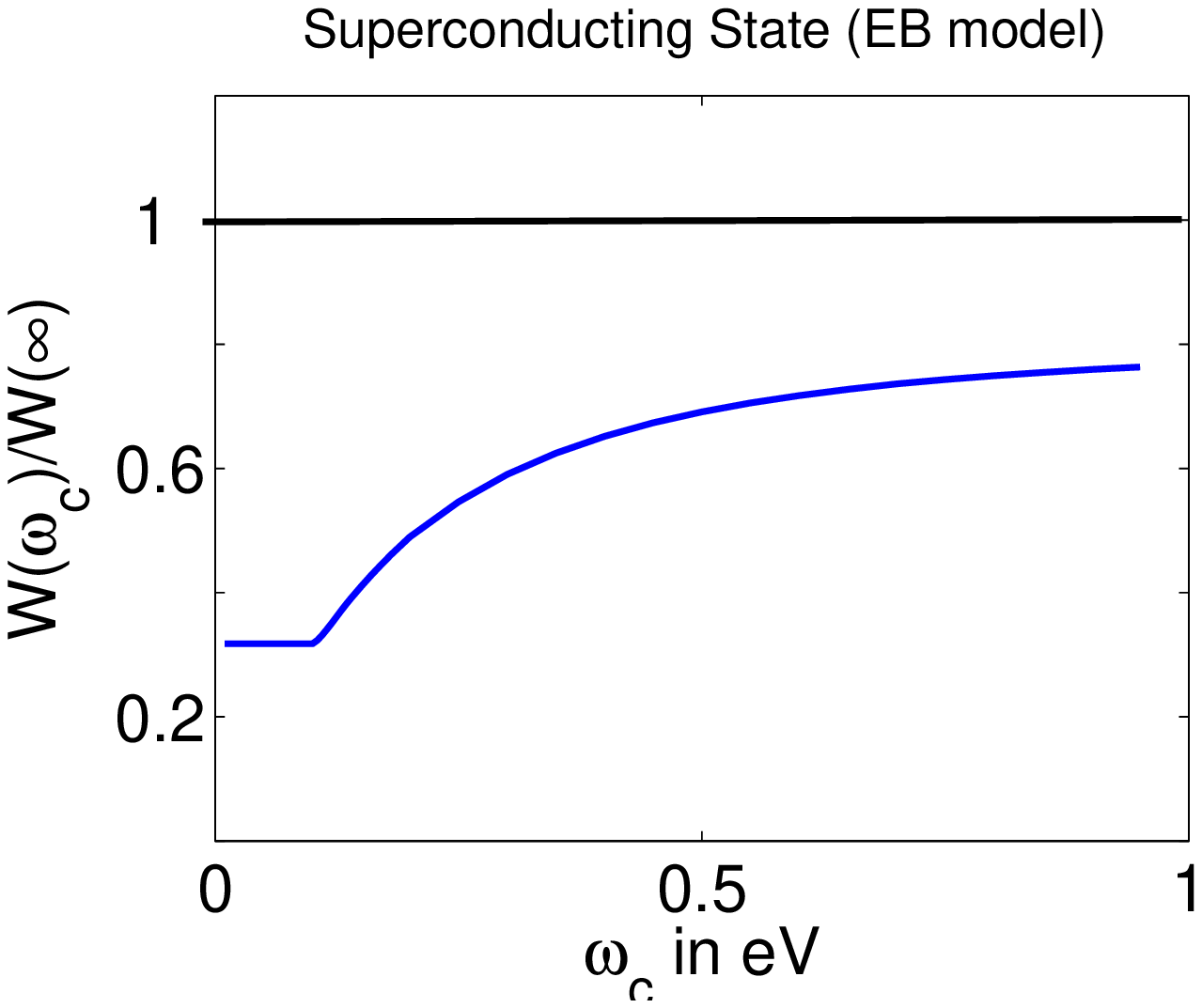}
\caption{\label{fig:PM_OS} Evolution of the optical integrals in the
EB model. Note that $W(0)$  has a non zero value at $T=0$ in the NS
because the self-energy at small frequencies is purely real and linear in $\omega$, hence the polarization bubble $\Pi (0) \neq 0$, as in an ideal Fermi gas.
 Parameters are the same as in fig. \ref{fig:PM_KS_Cond}}
\end{figure}

\begin{figure}[htp]
\includegraphics*[width=.7\linewidth]{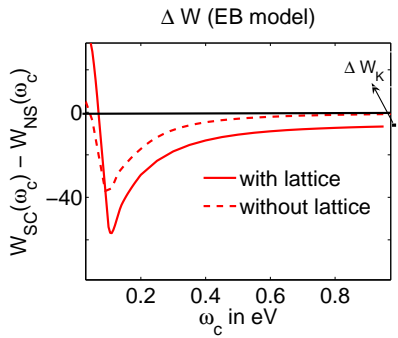}
\caption{\label{fig:PM_optdiff}$\Delta W$ vs the cut-off for the
EB model. It remains negative for larger cut-offs.
Parameters are the same as before. The dot indicates the value of $\Delta
W (\infty)=\Delta W_K$}
\end{figure}

Fig\ref{fig:PM_KS_Cond} shows conductivities $\sigma (\omega)$
 and Kubo sums $W_K$ as a function of the dimensionless
coupling $\lambda$. We see that, like in the previous case, the
Kubo sum in the NS is larger than that in the SCS. The difference
$\Delta W_K$ is between 5 and 8 meV.

 Fig \ref{fig:PM_OS} shows the evolution of the optical
integrals. Here we see the difference with the BCSI model --
 only about $75\%$ of the optical integral is recovered, both in the NS and SCS,
when we integrate up to the bandwidth of $1eV$. The rest comes from higher frequencies.

 In Fig \ref{fig:PM_optdiff} we plot $\Delta W (\w_c)$ as a function of $\w_c$. We see the same behavior as in the BCSI model in a clean limit -- $\Delta W (\w_c)$ is positive at small frequencies, crosses zero at some $\w_c$, passes through a deep minimum at a larger frequency, and eventually saturates at a negative value  at the largest $w_c$.  However, in distinction to BCSI model,
 $\Delta W (\w_c)$ keeps varying with $\w_c$ up a much larger scale and
 saturates only at around $0.8 eV$.  In between the dip at $0.1 eV$
and $0.8 eV$, the behavior of the optical integral is
predominantly determined by the variation of the cut-off term
$\Delta f (\w_c)$ as evidenced by a close similarity between the
behavior of the actual $\Delta W$ and $\Delta W$ in the absence of
the lattice (the dashed line in Fig. \ref{fig:PM_optdiff}).

%%%%%%%%%%%%%%%%%%%%%%%%%%%%%%%%%%%%%%%%%%%%%%%%%%%%%%%%%%%%%%%%%%%%%%%
\subsection{Marginal Fermi liquid model}\label{subsec: mode model}

For their analysis of the optical integral,
Norman and P\'{e}pin \cite{bib:norman pepin} introduced
 a  phenomenological model
 for the self energy which fits normal state scattering rate measurements by
ARPES\cite{bib:T valla}. It
constructs the NS $\Sigma^{''} (\omega)$  out of two contributions -
 impurity scattering and  electron-electron
scattering which they approximated phenomenologically by the
 marginal Fermi liquid form of $\alpha \omega$ at small frequencies~\cite{bib:MFL} (MFLI model). The total $\Sigma^{''}$  is
\begin{equation}
\label{eq:MM_ImSE}
\Sigma''(\omega)=\Gamma\,+\,\alpha |\omega| f\left(\frac{\omega}{\omega_{sat}}\right)
\end{equation}
where $\omega_{sat}$ is about $\thicksim\frac{1}{2}$ of the
bandwidth, and $f(x) \approx 1$ for $x <1$ and decreases for $x
>1$. In Ref \onlinecite{bib:norman pepin} $f(x)$ was assumed to scale as
$1/x$ at large $x$ such that $\Sigma''$ is flat at large $\omega$.
The real part of $\Sigma (\omega)$ is obtained from
Kramers-Kr\"{o}nig relations. For the superconducting state, they
obtained $\Sigma^{''}$ by cutting off the NS expression on the
lower end at some frequency $\omega_1$ (the analog of $\omega_0 +
\Delta$ that we had for EB model):
\begin{equation}
\label{eq:MM_ImSE_SC} \Sigma''(\omega)=(\Gamma\,+\,\alpha
|\omega|)\Theta(|\omega|-\omega_1)
\end{equation}
where $\Theta(x)$ is the step function.
 In reality, $\Sigma^{''}$ which fits ARPES in the NS has some angular dependence along the Fermi surface \cite{bib:ARPES_anisotropy}, but this was ignored
 for simplicity. This model had gained a lot of attention as it
predicted the optical sum in the SCS to be larger than in the NS, i.e.,
 $\Delta W>0$ at large frequencies.  This would be consistent with the
experimental findings in Refs. \onlinecite{bib:molegraaf,
bib:optical int expt} if, indeed,
 one identifies  $\Delta W$ measured up to 1eV with $\Delta W_K$.

\begin{figure}[htp]
\includegraphics*[width=.7\linewidth]{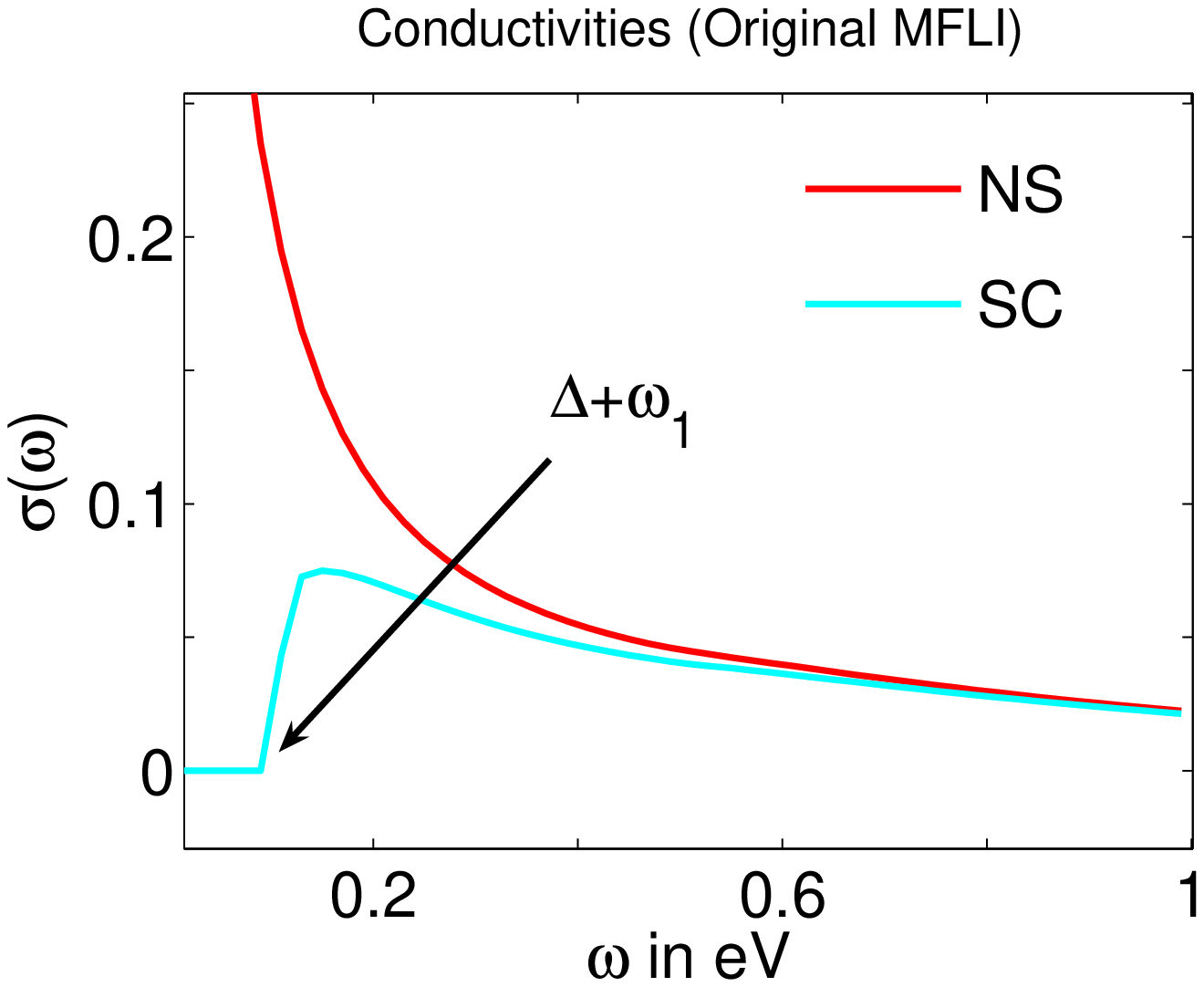}
\hfill
\includegraphics*[width=0.7\linewidth]{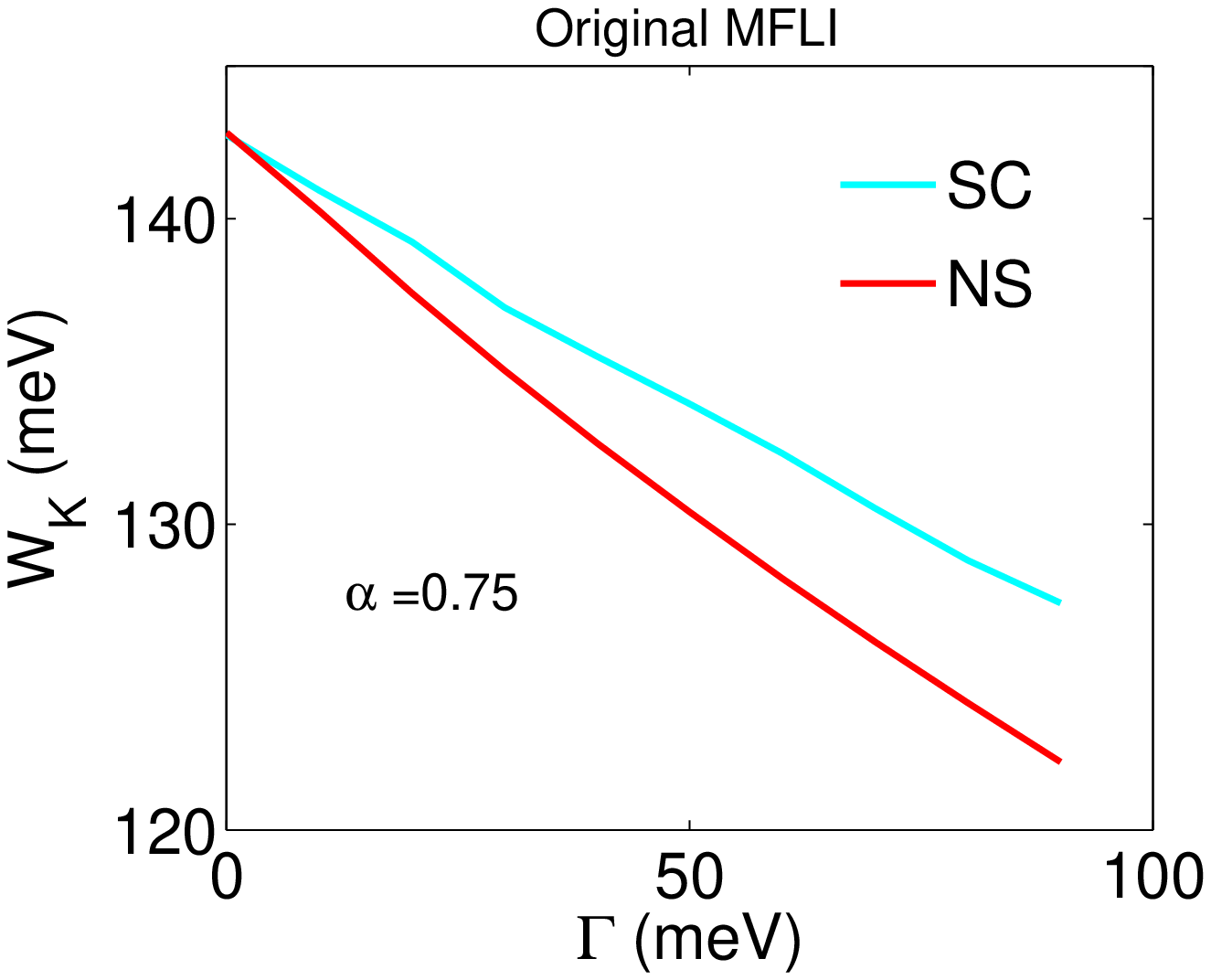}
\caption{\label{fig:MM_KS_cond}Top --the conductivities in the
 NS and SCS in
the original  MFLI model of Ref.\onlinecite{bib:norman pepin}. We
set
 $\Gamma = 70\,meV$, $\alpha=0.75$,
$\Delta=32\,meV$, $\omega_1=71\,meV$. Note that $\sigma' (\omega)$ in the SCS
begins at $\Omega=\Delta+\omega_1$. Bottom -- the behavior of $W_K$ with
$\Gamma$.}
\end{figure}

\begin{figure}[htp]
\includegraphics*[width=.7\linewidth]{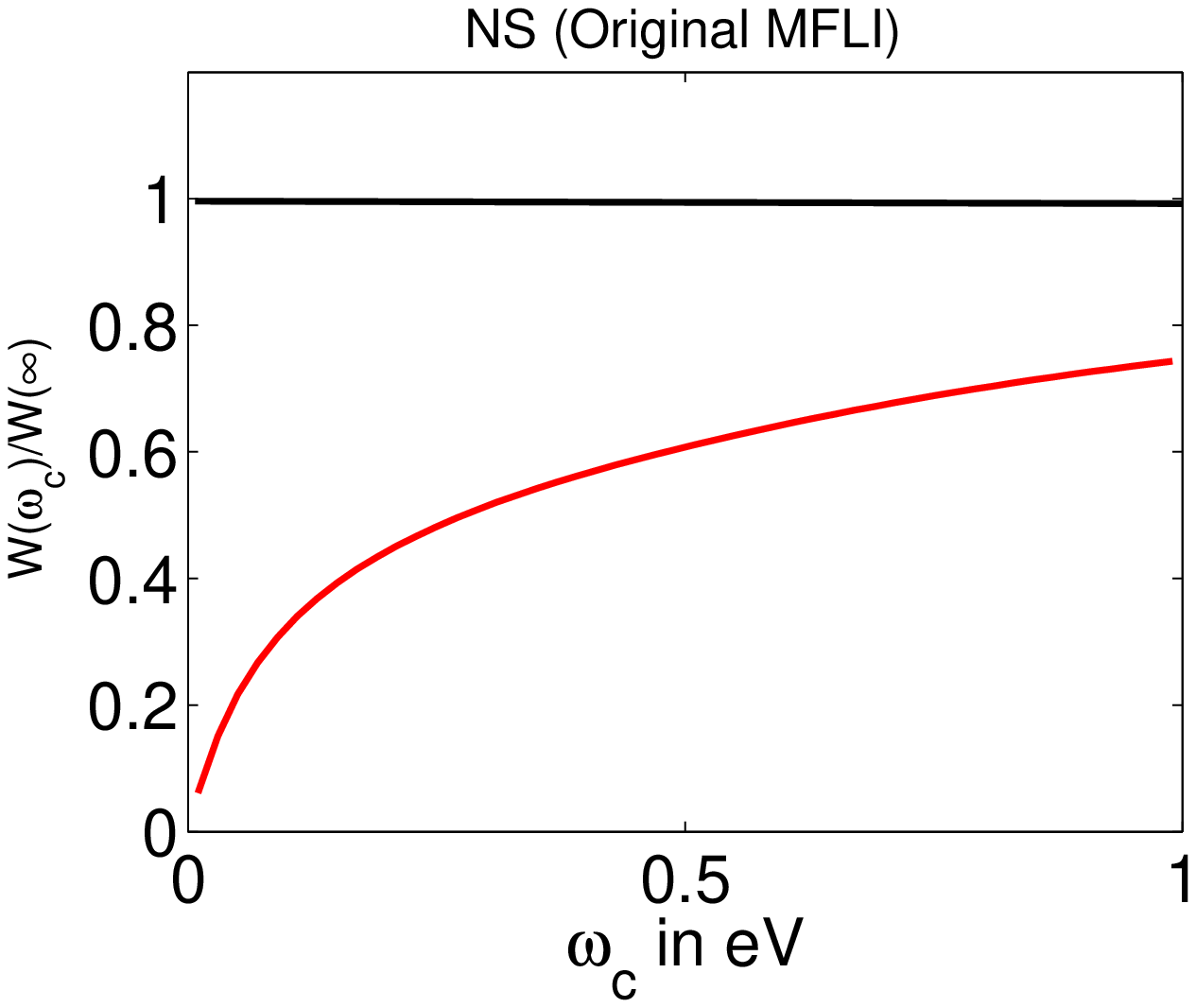}
\hfill
\includegraphics*[width=.7\linewidth]{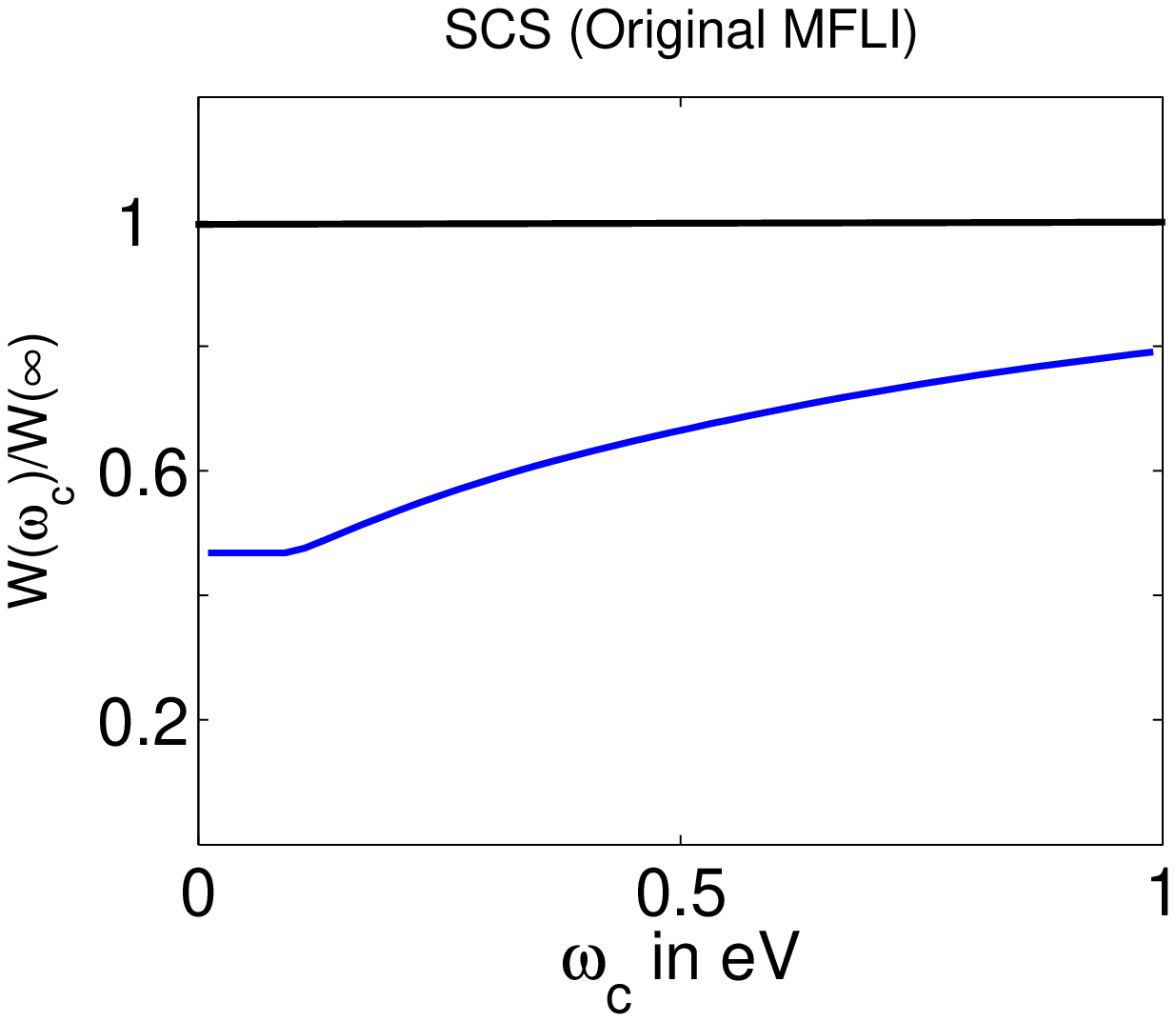}
\caption{\label{fig:MM_OS} The evolution of the optical integral in
the NS (top) and the SCS (bottom) in the original MFLI model.
 Parameters are the same as
above. Note that only $\sim 75-80\%$ of the spectral weight
 is recovered up to $1 eV$.}
\end{figure}

\begin{figure}[htp]
\includegraphics*[width=.7\linewidth]{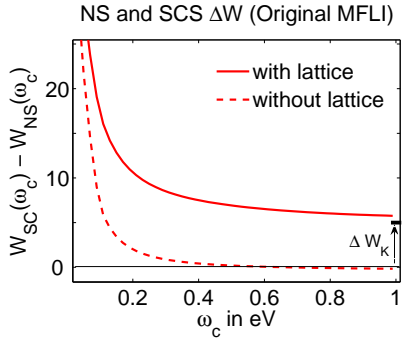}
\caption{\label{fig:MM_OpDf}Evolution of the difference of the optical
integrals in the SCS and the NS with the upper cut-off $\omega_c$.
Parameters are the same as before. Observe that the optical sum in the SCS is
larger than in the NS and that $\Delta W$ has not yet reached $\Delta
W_K$ up to the bandwidth. The dashed line is the FGT result.}
\end{figure}

 We will show below that the sign of $\Delta W$ in the MFLI model actually
 depends on how the normal state results are extended to the superconducting state and, moreover, will argue that $\Delta W_K$ is actually negative if the extension is done such that at $\alpha =0$ the results are consistent with BCSI model.
  However, before that, we show in Figs
\ref{fig:MM_KS_cond}-\ref{fig:MM_OpDf} the conductivities and the optical integrals for the original MFLI model.

\begin{figure}[htp]
\includegraphics*[width=.7\linewidth]{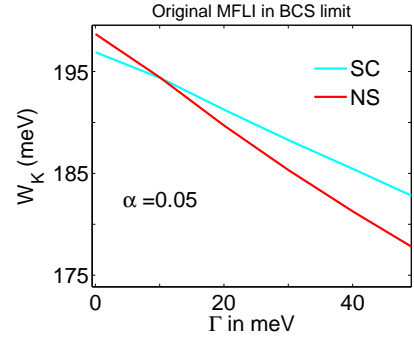}
\caption{\label{fig:MM_sp_KS} Behavior of $W_K$ with $\Gamma$ for
the original MFLI model at very small $\alpha=0.05$. We set
$\omega_1=\Delta=32\,meV$. Observe the inconsistency with $W_K$ in
the BCSI model in Fig \ref{fig:BCS_KS_Cond}.}
\end{figure}

\begin{figure}[htp]
\includegraphics*[width=.7\linewidth]{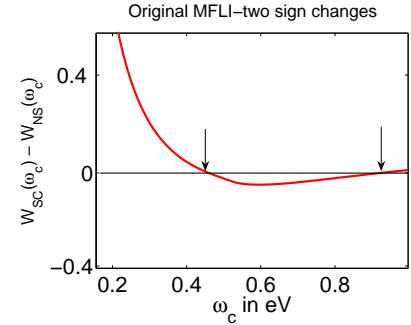}
\caption{\label{fig:MM_optdiff_2sgnchng} The
special case of $\alpha=1.5$,$\Gamma=5\,meV$, other parameters the same as in Fig. \protect\ref{fig:MM_KS_cond}. These parameters are chosen to illustrate
 that two sign changes (indicated by arrows in the figure) are also
possible within the original MFLI model.}
\end{figure}

\begin{figure}[htp]
\includegraphics*[width=.7\linewidth]{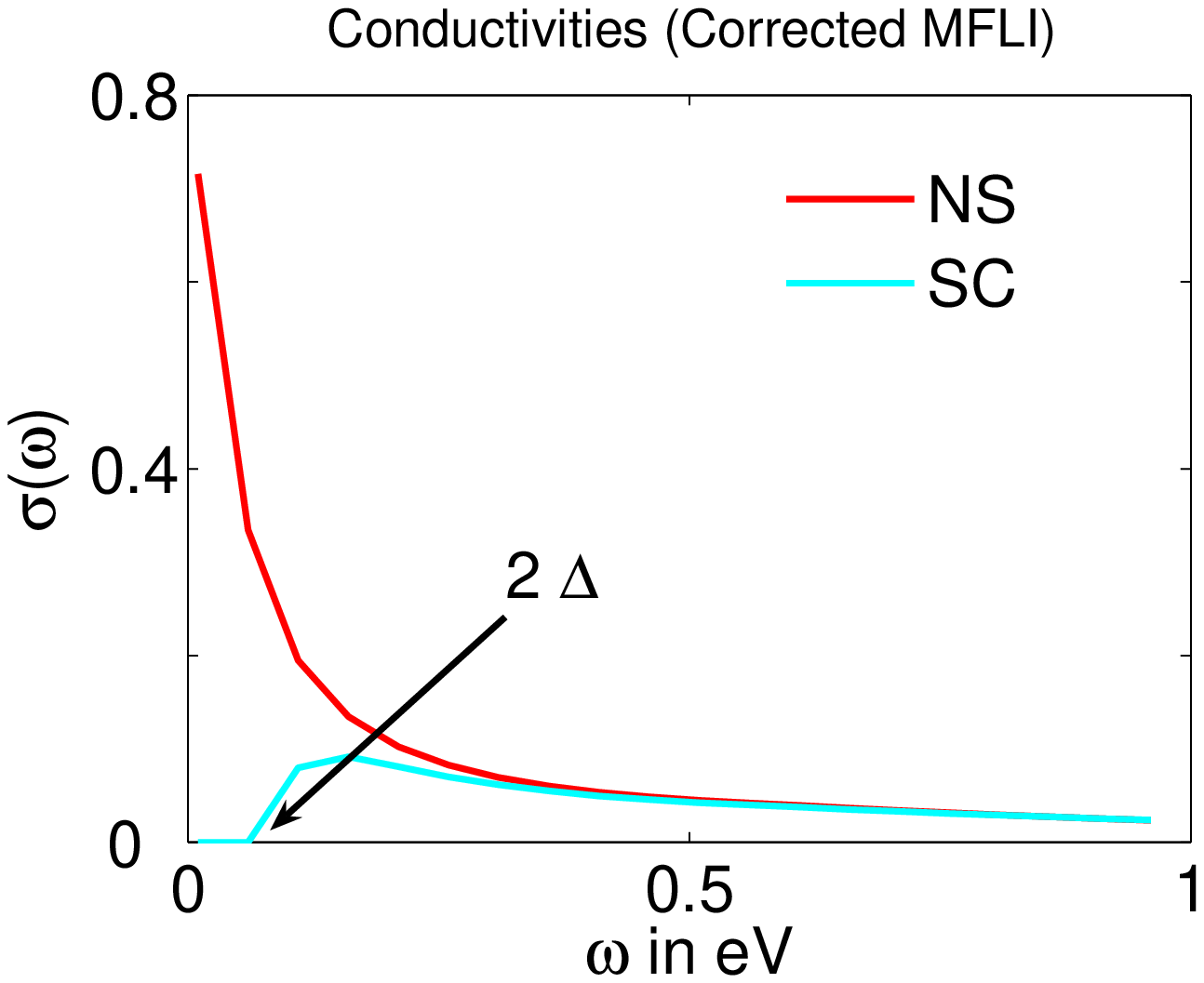}
\hfill
\includegraphics*[width=.7\linewidth]{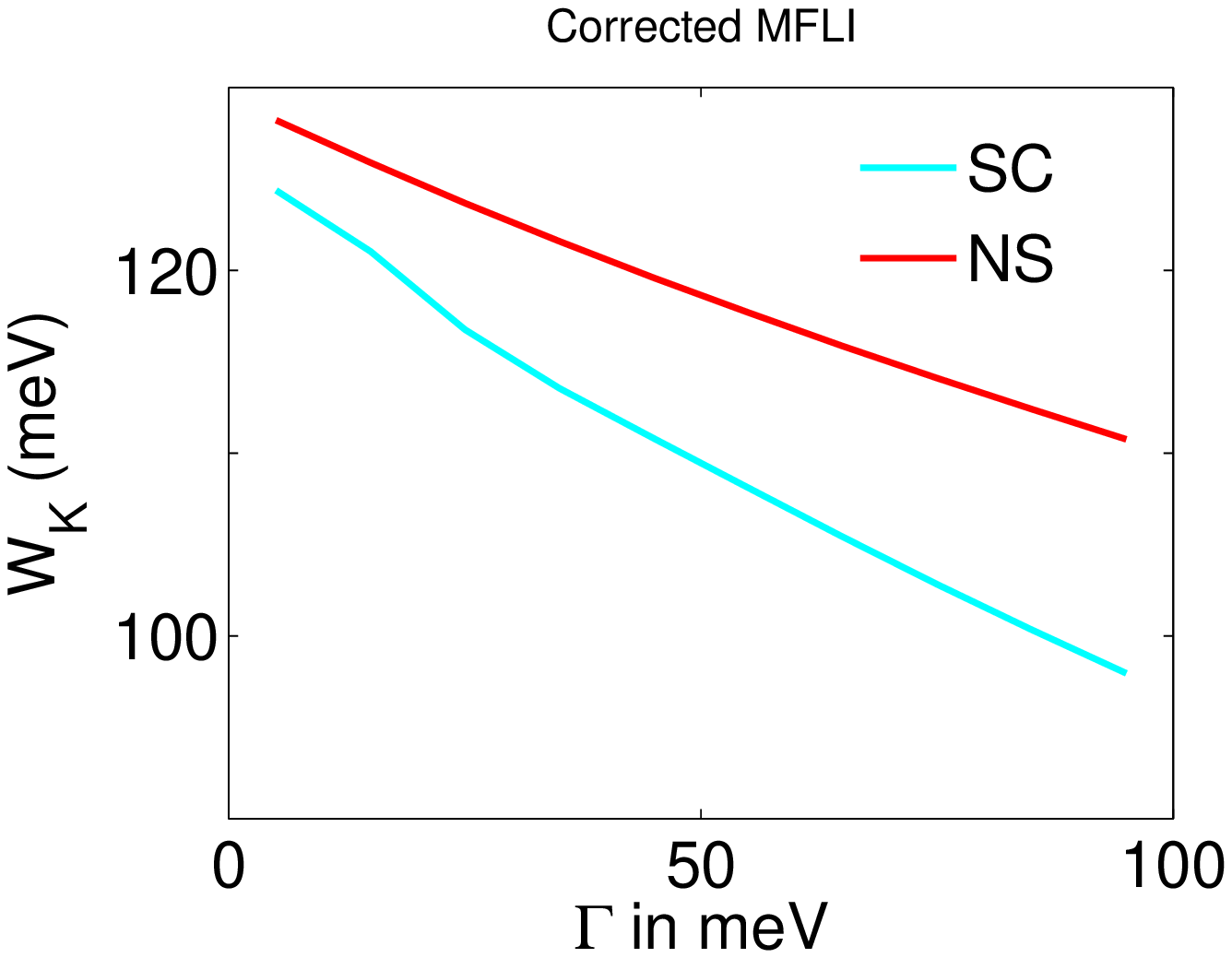}
\caption{\label{fig:MM_corr_KS_cond}Top -- $\sigma (\omega)$ in the NS and the
SCS in the `corrected' MFLI model with the feedback from SC on the quasiparticle damping: $i\Gamma$ term transforms into
$\frac{\Gamma}{\sqrt{-\omega^2+\Delta^2}}$.  In the SCS $\sigma$ now begins at
$\Omega=2\Delta$.
The parameters are same as in  Fig. \protect\ref{fig:MM_KS_cond}.
Bottom -- the  behavior of Kubo sum
with $\Gamma$. Observe that $W(\w_c)$ in the NS  is larger than in the SCS.}
\end{figure}

\begin{figure}[htp]
\includegraphics*[width=.7\linewidth]{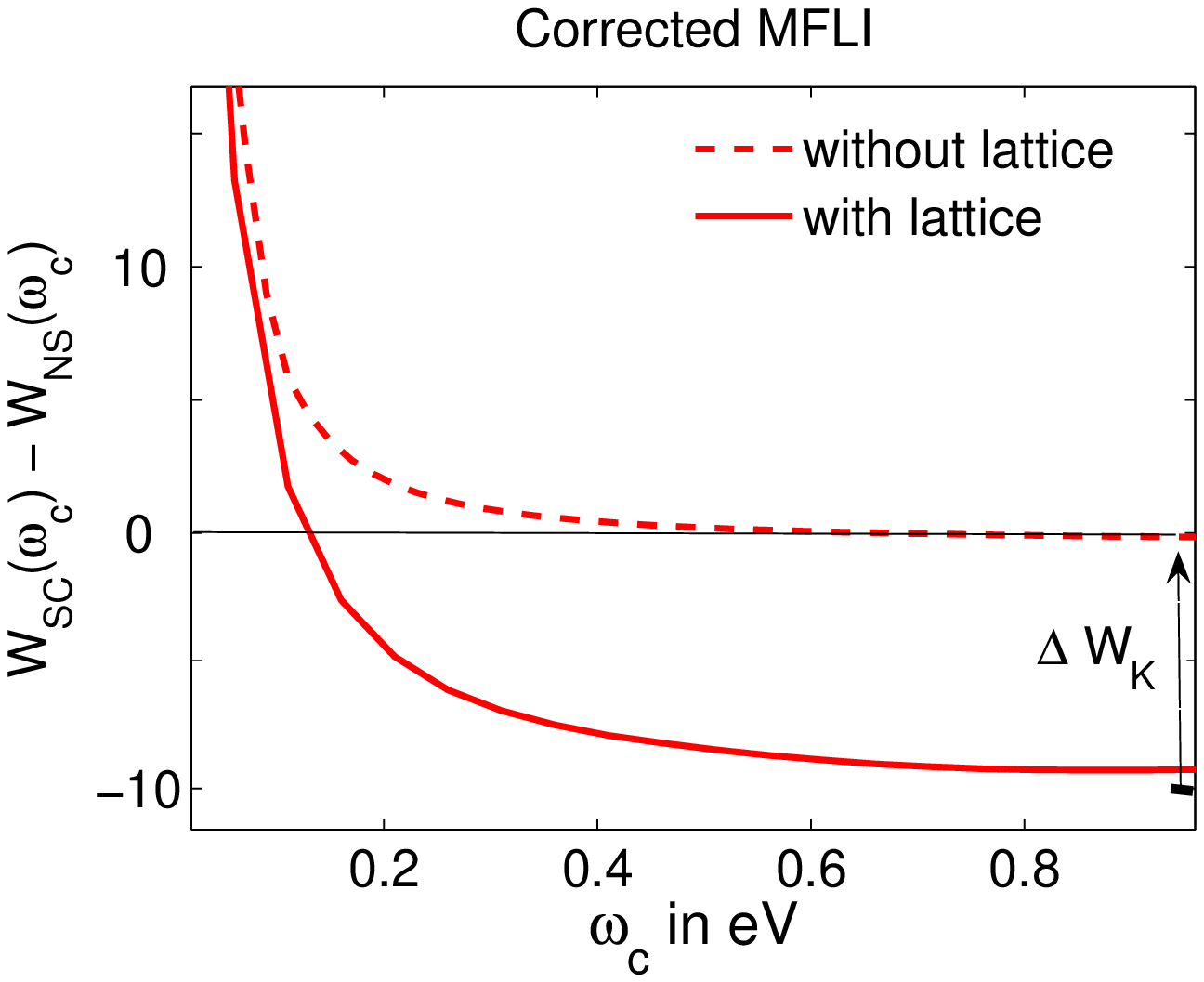}
\caption{\label{fig:MM_corr_OpDf}Evolution of the difference of
the optical integrals between the SCS and the NS with the upper cut-off
$\omega_c$ for the ``corrected'' MFLI model. Now $\Delta W(\w_c)$ is negative above some frequency. Parameters are same as in the
Fig \ref{fig:MM_corr_KS_cond}.}
\end{figure}

In Fig \ref{fig:MM_KS_cond} we plot the conductivities in the NS
and the SCS and Kubo sums $W_K$ vs $\Gamma$ at $\alpha =0.75$
showing that the spectral
weight in the SCS is indeed larger than in the NS.
In Fig \ref{fig:MM_OS} we show the behavior of the
optical sums $W(\w_c)$ in NS and SCS. The observation here is that only
 $\sim75-80\%$ of the Kubo sum is recovered up to the
scale of the bandwidth implying that there is indeed a significant
spectral weight well beyond the bandwidth. And in Fig
\ref{fig:MM_OpDf} we show the behavior of $\Delta W (w_c)$. We see that it
 does not change sign and remain positive at all $\w_c$,
 very much unlike the BCS case. Comparing the behavior of $W(w_c)$ with and without a lattice (solid and dashed lines in Fig. \ref{fig:MM_OpDf})  we see that
 the `finite bandwidth effect' just  shifts the curve in the positive
direction. We also see that the solid line flattens above roughly half of the bandwidth, i.e., at these frequencies $\Delta W(\w_c) \approx \Delta W_K$.
 Still, we found that $\Delta W$
 continues going down even above the bandwidth and truly saturates only at
  about $2\,eV$ (not shown in the figure) supporting the idea that there is `more' left to recover from
  higher frequencies.

The rationale for $\Delta W_K >0$ in the original MFLI model has
been provided
 in Ref.~\onlinecite{bib:norman pepin}. They argued that this  is closely linked
to the absence of quasiparticle peaks in the NS and their restoration in the SCS state because the phase space for quasiparticle scattering at low energies is smaller in a superconductor than in a normal state. This clearly affects $n_k$
 because it is expressed via the full Green's function and competes with the
 conventional effect of the gap opening.
 The distribution function from this model, which we show in
 Fig.\ref{fig:Dist fns}b
  brings this point out by showing that in a MFLI model, at $\epsilon<0$,
 $n_k$  in a superconductor is larger than $n_k$ in the normal state,
 in clear difference with the BCSI case.

 We analyzed the original MFLI model for various parameters and found  that
 the behavior presented in Fig. \ref{fig:MM_OpDf}, where $\Delta W (\w_c)>0$ for all frequencies,  is typical but not not a generic one. There exists a  range
 of  parameters $\alpha$ and  $\Gamma$ where $\Delta W_K$ is still positive, but $\Delta W (\w_c)$ changes the sign twice and is negative at intermediate frequencies. We show an example of  such behavior in
 Fig\ref{fig:MM_optdiff_2sgnchng}.  Still, for most of the parameters, the behavior of $\Delta W(\w_c)$ is the same as in  Fig. \ref{fig:MM_OpDf}.

On more careful looking we  found the problem with the original
MFLI model. We recall that in this model the self-energy in the
 SCS state was obtained by just cutting the NS
self energy at $\omega_1$ (see Eq.\ref{eq:MM_ImSE_SC}). We argue
that
 this phenomenological formalism is not fully consistent, at least for
 small $\alpha$.  Indeed, for $\alpha =0$, the MFLI model reduces to BCSI model
 for which
 the behavior of the self-energy is given by Eq. (\ref{eq:sigma_bcs}).
This self-energy evolves with $\omega$ and $\Sigma^{''}$ has
 a square-root singularity at $\omega = \Delta + \omega_o$ (with $\omega_o = 0$). Meanwhile
 $\Sigma^{''}$  in the original MFLI model in Eq. (\ref{eq:MM_ImSE_SC})
 simply jumps to zero at $\omega = \omega_1 =\Delta$, and this happens for all values
 of $\alpha$ including $\alpha =0$ where the MFLI and BCSI model should merge.
 This inconsistency is reflected in Fig \ref{fig:MM_sp_KS}, where we
plot the near-BCS limit of MFLI model by taking a very small
$\alpha=0.05$. We see that the optical integral $W_K$ in the SCS
still remains larger than in the NS over a wide range of $\Gamma$,
in clear difference with the exactly known  behavior in the BCSI
model, where $W_K$ is larger in the NS for all $\Gamma$ (see Fig.
\ref{fig:BCS_KS_Cond}). In other words, the original MFLI model
does not have the BCSI theory as its limiting case.

We modified the MFLI model is a minimal way by changing the damping term in
a SCS to $\frac{\Gamma}{\sqrt{-\omega^2+\Delta^2}}$ to be consistent with BCSI model. We still use Eq. (\ref{eq:MM_ImSE_SC}) for the MFL term simply because this term was introduced in the NS on phenomenological grounds and there is no way to guess how it gets modified in the SCS state without first deriving
 the normal state self-energy  microscopically (this is what we will do in the next section). The results of the calculations for the modified MFLI model are presented in Figs. \ref{fig:MM_corr_KS_cond} and
\ref{fig:MM_corr_OpDf}. We clearly see that the behavior is now different and $\Delta W_K < 0$ for all $\Gamma$. This is the same behavior as we previously found  in BCSI and EB models.
 So we argue that the `unconventional' behavior exhibited by the original MFLI
 model is most likely the manifestation of a particular modeling inconsistency.
Still, Ref. \onlinecite{bib:norman pepin} made a valid point that the fact that quasiparticles behave more close to free fermions in a SCS than in a
 NS, and this effect tends to reverse the signs of $\Delta W_K$ and of the
 kinetic energy ~\cite{bib:haslinger}. It just happens that in a modified MFLI model the optical integral is still larger in the NS.

%%%%%%%%%%%%%%%%%%%%%%%%%%%%%%%%%%%%%%%%%%%%%%%%%%%%%%%%%%%%%%%%%%%%%%
\subsection{The collective boson model}

 We now turn to a more microscopic  model- the CB model.
The model describes fermions
interacting by exchanging soft, overdamped collective bosons in a particular, near-critical, spin or charge channel~\cite{bib:acs,bib:fink,bib:GRILLI}.
 This interaction is responsible for the normal state self-energy and also gives rise to a superconductivity.
 A peculiar feature of the CB model is that the propagator of
 a collective boson changes below $T_c$ because this boson is not an independent degree of freedom (as in EB model) but is made out of low-energy fermions which are affected by superconductivity~\cite{bib:coupling to bos mode}.

The most relevant
 point for our discussion is that this model contains the physics which we
 identified above as a source of a potential
 sign change of $\Delta W_K$. Namely, at strong coupling
 the fermionic self-energy in the NS is large because there exists
 strong scattering between low-energy fermions mediated by low-energy collective bosons. In the SCS, the density of low-energy fermions drops and a continuum  collective excitations becomes gaped. Both effects reduce fermionic damping and
 lead to the increase of $W_K$ in a SCS. If this increase exceeds a conventional loss of $W_K$ due to a gap opening, the total $\Delta W_K$ may become positive.

The CB model has been applied numerous times to the cuprates, most
often under the assumption that near-critical collective
excitations are spin fluctuations with momenta near $Q =
(\pi,\pi)$. This version of a CB boson is commonly known as a
spin-fermion model. This model yields $d_{x^2-y^2}$
superconductivity and explains in a quantitative way a number of
measured electronic features of the cuprates, in particular the
near-absence of the quasiparticle peak in the NS of optimally
doped and underdoped cuprates\cite{bib:no quasi NS} and the
peak-dip-hump structure in the ARPES profile in the
SCS\cite{bib:peak dip hump, bib:coupling to bos mode, bib:fink,
bib:coll modes mike ding}. In our analysis we assume that a CB is
a spin fluctuation.

The results for the conductivity within a spin-fermion model
depend in quantitative (but not qualitative) way on the assumption
for the momentum dispersion of a collective boson. This momentum
dependence comes from high-energy fermions and is an input for the
low-energy theory. Below we follow Refs. \onlinecite{bib:sum rule
mike_chu, bib:fink} and assume that the momentum dependence of a
collective boson is flat near $(\pi,\pi)$. The self energy within
such model has been worked out consistently in Ref.
\onlinecite{bib:sum rule mike_chu, bib:fink}. In the normal state
\[
\Sigma''(\omega)=-\frac{1}{2}\,\lambda_n\omega_{sf}\, log\left(
1+\frac{\omega^2}{\omega_{sf}^2}\right)
\]
\begin{equation}
\Sigma'(\omega)= -\lambda_n  \omega_{sf} \,
arctan\frac{\omega}{\omega_{sf}}
\end{equation}
where $\lambda_n$ is the spin-fermion coupling constant, and
 $\omega_{sf}$ is a typical spin relaxation frequency of overdamped spin collective excitations with a propagator
\begin{equation}
\label{pl_1}
\chi (q \sim Q, \Omega) = \frac{\chi_Q}{1 - i
\frac{\Omega}{\omega_{sf}}}
\end{equation}
where $\chi_Q$ is the uniform static susceptibility. If we use
Ornstein-Zernike form of $\chi (q)$ and use either Eliashberg
~\cite{bib:acs} or  FLEX computational schemes~\cite{bib:FLEX}, we
get rather  similar behavior of $\Sigma$ as a function of
frequency and rather similar behavior of optical integrals.

The collective nature of spin fluctuations is reflected in the
fact that the coupling $\lambda$ and the bosonic frequency
$\omega_{sf}$ are related: $\lambda$ scales as $\xi^2$, where
$\xi$ is the bosonic mass (the distance to a bosonic instability),
and $\omega_{sf} \propto \xi^{-2}$ (see Ref. \onlinecite{bib:disp
anamoly}). For a flat $\chi (q \sim Q)$ the  product $\lambda
\omega_{sf}$ does not depend on $\xi$ and is the overall
dimensional scale for boson-mediated interactions.

In the SCS fermionic excitations acquire a gap. This gap affects
fermionic self-energy in two ways: directly, via the change of the
dispersion of an intermediate boson in the exchange process
involving a  CB, and indirectly, via the change of the propagator
of a CB. We remind ourselves that the dynamics of a CB comes from
a particle-hole bubble which is indeed affected by $\Delta$.

The effect of a $d-$wave pairing gap on a CB has been discussed in
a number of papers, most recently in~\cite{bib:fink}.
 In a SCS a gapless continuum described by Eq.
(\ref{pl_1}) transforms into a gaped continuum, with a gap about
$2\Delta$ and a resonance at $\omega = \omega_0 < 2\Delta$, where
for a $d-$wave gap we define $\Delta$ as a maximum of a $d-$wave
gap.

The spin susceptibility near $(\pi,\pi)$  in a superconductor can generally be written up as
\begin{equation}\label{eq:chi}
\chi (q \sim Q, \Omega) = \frac{\chi_Q}{1 - i \frac{\Pi
(\Omega)}{\omega_{sf}}}
\end{equation}
where $\Pi$ is evaluated by adding up the bubbles made out of two
normal and two anomalous Green's functions. Below $2\Delta$, $\Pi
(\Omega)$ is real ($\sim \Omega^2/\Delta$ for small $\Omega$), and
the resonance emerges at $\Omega = \omega_0$ at which $\Pi
(\omega_0) = \omega_{sf}$. At frequencies larger than $2\Delta$,
$\Pi (\Omega)$ has an imaginary part, and this gives rise to a
gaped continuum in $\chi (\Omega)$.

The imaginary part of the spin susceptibility around
the resonance frequency $\omega_0$ is~\cite{bib:fink}
\begin{equation}\label{eq:SCS_chi}
\chi^{''}(q, \Omega)= \frac{\pi Z_o \omega_0}{2} \delta (\Omega - \omega_0)
%\frac{\omega_o^2}{\omega_o^2-\Omega^2-i\delta}
\end{equation}

where $Z_o \sim 2\,\omega_{sf} \chi_0/\frac{\partial \Pi}{\partial
\omega}_{|\Omega = \omega_0}$ . The imaginary part of the
spin susceptibility describing a gaped
continuum exists for for $\Omega \geq 2\Delta$ and is
\[\chi^{''}(q, \Omega)= Im \left[\frac{\chi_0}{1-\frac{1}{\omega_{sf}}\left(\frac{4
\Delta^2}{\Omega}D(\frac{4 \Delta^2}{\Omega^2})+i\Omega
K_2(1-\frac{4 \Delta^2}{\Omega^2})\right)}\right]
\]
\begin{equation}\label{eq:SCS_chi_continiuum}
\approx Im\left[\frac{\chi_0}{1-\frac{1}{\omega_{sf}}\left(\frac{\pi
\Delta^2}{\Omega}+i\frac{\pi}{2}\Omega \right)} \right] \; {\text for}\; \Omega>>2\Delta
\end{equation}
In Eq. (\ref{eq:SCS_chi_continiuum})
 $D(x)=\frac{K_1(x)-K_2(x)}{x}$, and $K_1(x)$ and $K_2(x)$
are Elliptic integrals of first and second kind.
The real part of $\chi$ is obtained by Kramers-Kr\"{o}nig
transform of the imaginary part.

Substituting  Eq \ref{eq:NS_SCS_Self} for $\chi (q, \Omega)$ into
the formula for the self-energy one obtains $\Sigma'' (\omega)$ in
a SCS state as a sum of two terms~\cite{bib:fink}
\begin{widetext}
\begin{equation}
\Sigma''(\omega)=\Sigma''_A(\omega)+\Sigma''_B(\omega)
\end{equation}
where,
\[
\Sigma''_A(\omega)=\frac{\pi Z_o}{2}\,\lambda_n\omega_o\,Re \left(
\frac{\omega+\omega_o}{\sqrt{(\omega+\omega_o)^2-\Delta^2}}\right)
\]
comes from the interaction with the resonance and
\begin{equation}
\Sigma''_B(\omega)=-\lambda_n \int_{2\Delta}^{|E|}dx\,Re\,
\frac{\omega+x}{\sqrt{(\omega+x)^2-\Delta^2}} \frac{
\frac{x}{\omega_{sf}}K_2\left(1-\frac{4 \Delta^2}{x^2}\right)
}{\left[ 1-
\frac{4\Delta^2}{x\omega_{sf}}D\left(\frac{4\Delta^2}{x^2}\right)
\right]^2 + \left[ \frac{x}{\omega_{sf}}K_2\left(1-\frac{4
\Delta^2}{x^2}\right) \right]^2}
\label{pl_2}
\end{equation}
\end{widetext}
comes from the interaction with the gaped continuum.
The real part of $\Sigma$ is obtained by Kramers-Kr\"{o}nig
transform of the imaginary part.

\begin{figure}[htp]
\includegraphics*[width=.7\linewidth]{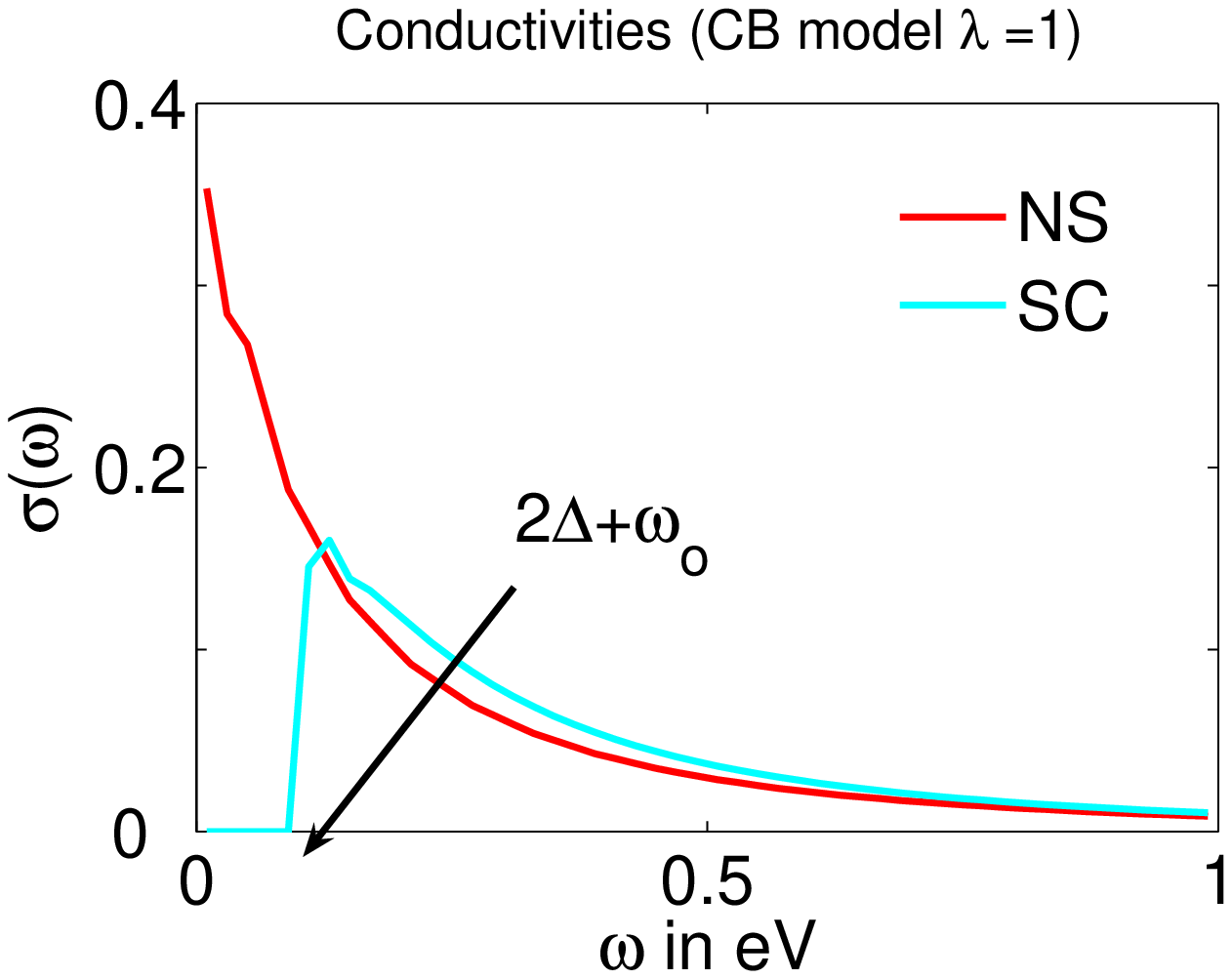}
\hfill
\includegraphics*[width=.7\linewidth]{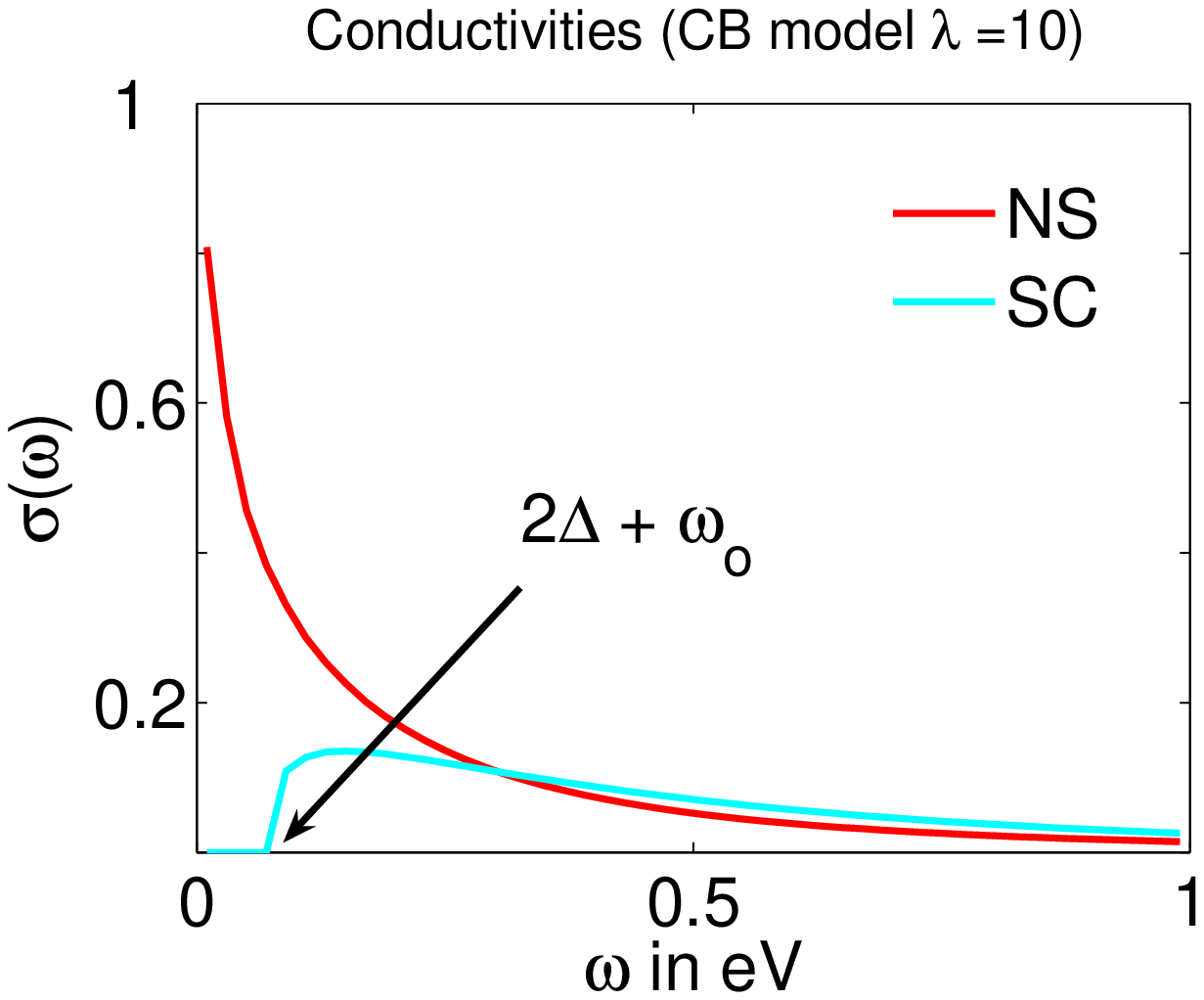}
\caption{\label{fig:CM_cond} Conductivities and $\Delta W$ for a fixed
 $\lambda \omega_{sf}$.
 Top --
$\omega_{sf}=26\,meV$,$\lambda=1$,$\omega_o=40\,meV$,$Z_o=0.77$
Bottom --
$\omega_{sf}=2.6\,meV$,$\lambda=10$,$\omega_o=13.5\,meV$,$Z_o=1.22$.
The zero crossing for $\Delta W$ is not affected by a change in
$\lambda$ because it is determined only by $\lambda\omega_{sf}$.
We set
 $\Delta = 30\,meV$.}
\end{figure}

\begin{figure}[htp]
\includegraphics*[width=.7\linewidth]{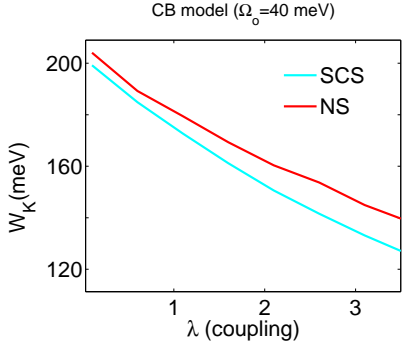}
\caption{\label{fig:CM_Kubosums_mode40}The behavior of Kubo sums
in the CB model. Note that the spectral weight in the
NS is always larger than in the SCS.
We set $\omega_{sf}=26\,meV$,$\lambda=1$, and  $\Delta = 30\,meV$.}
\end{figure}

\begin{figure}[htp]
\includegraphics*[width=.7\linewidth]{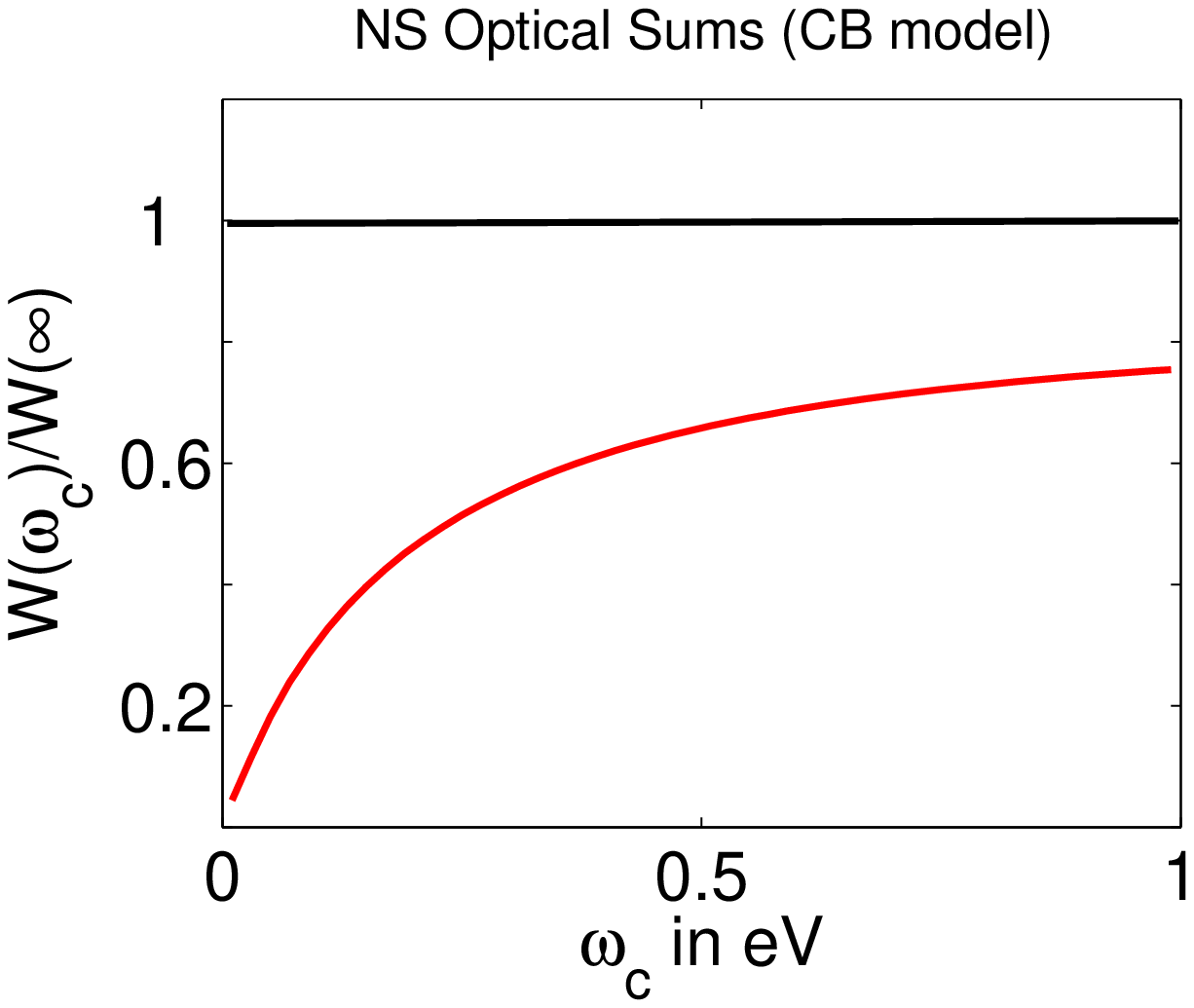}
\hfill
\includegraphics*[width=.7\linewidth]{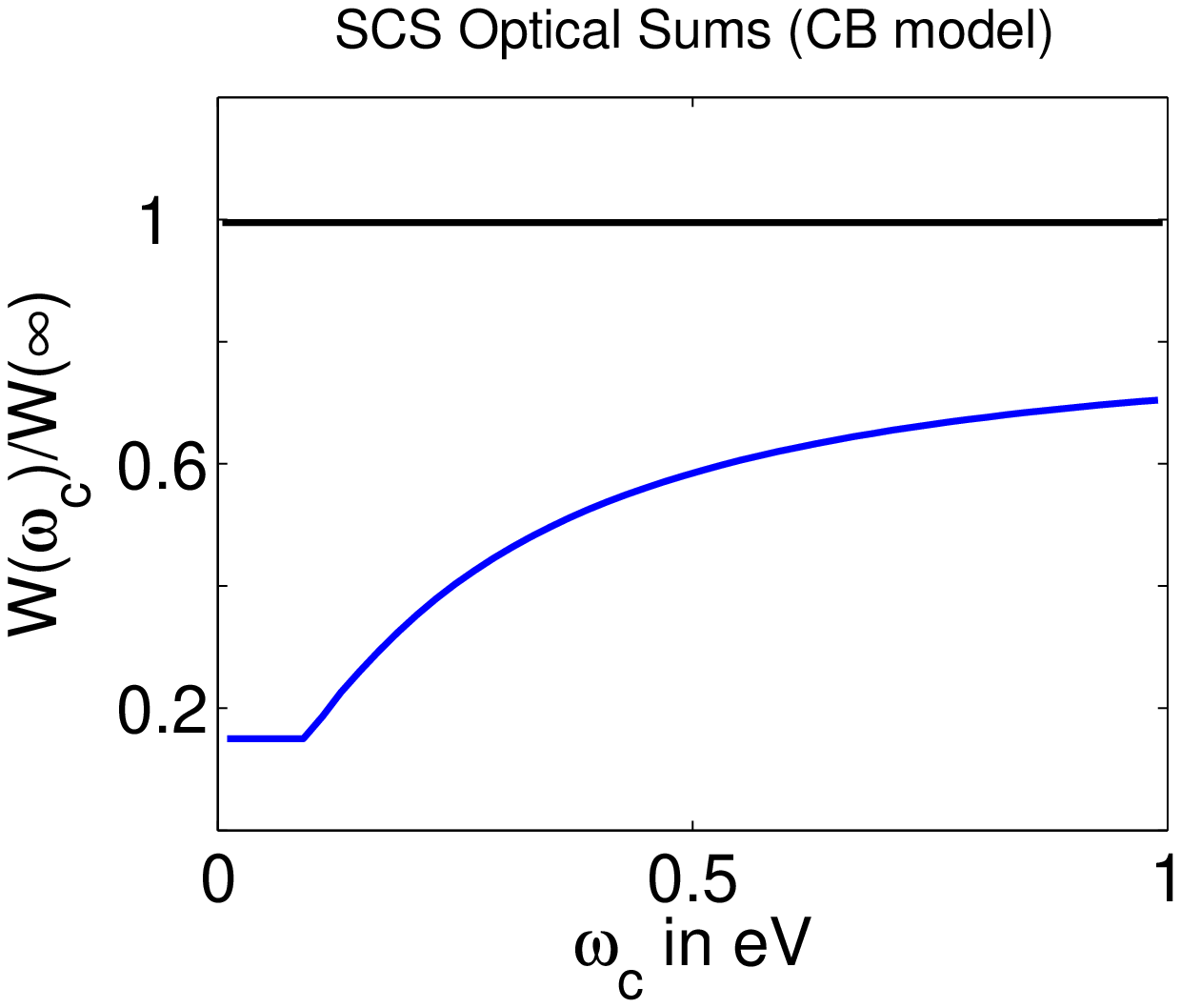}
\caption{\label{fig:CM_OS} The evolution of the optical integrals in
the NS and the SCS in the CB model.
Note that about  $\sim75\%$ of the spectral weight is recovered
 up to $1\,eV$.
We set $\omega_{sf}=26\,meV$,$\lambda=1$, and  $\Delta = 30\,meV$.}
\end{figure}

\begin{figure}[htp]
\includegraphics*[width=.7\linewidth]{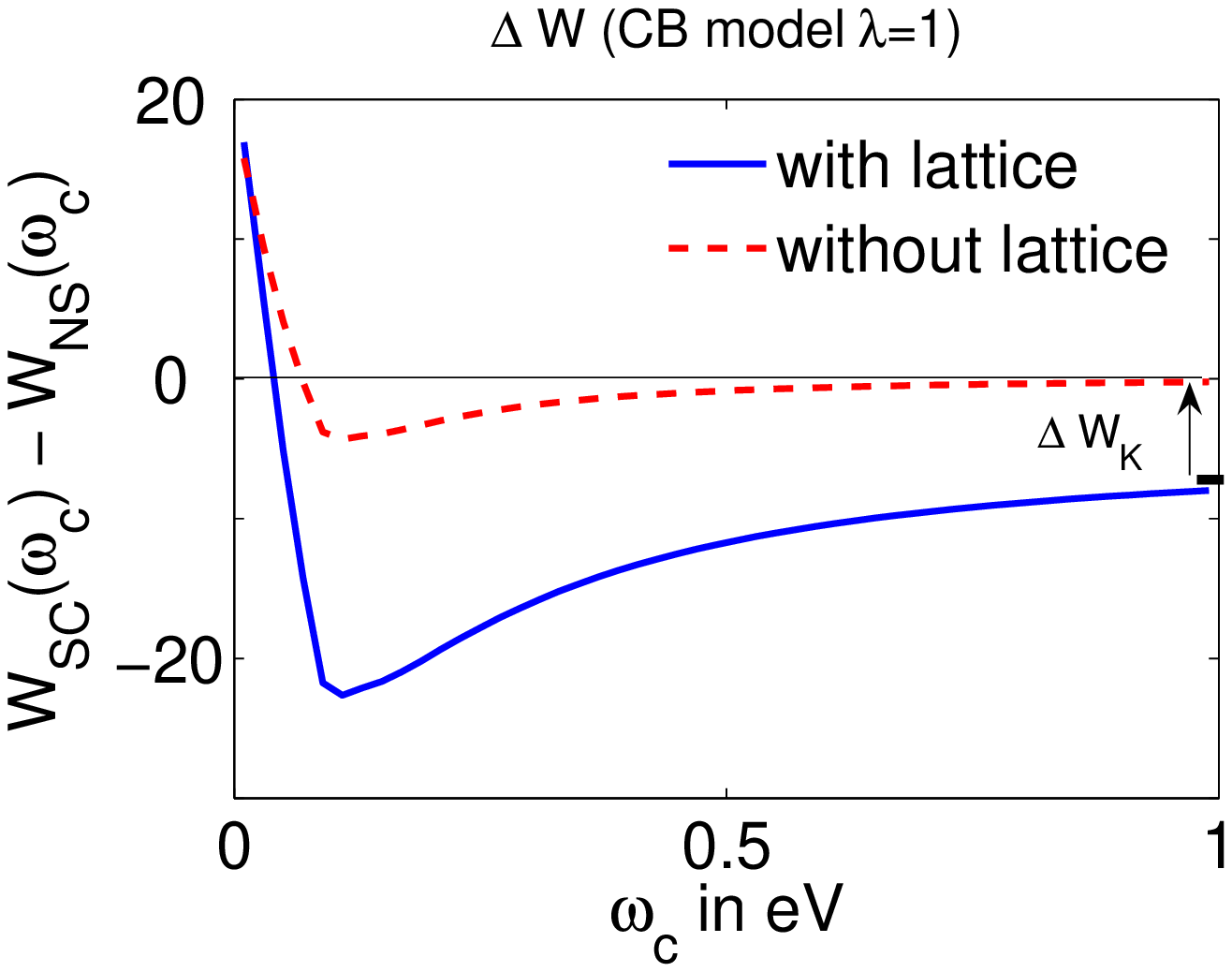}
\hfill
\includegraphics*[width=.7\linewidth]{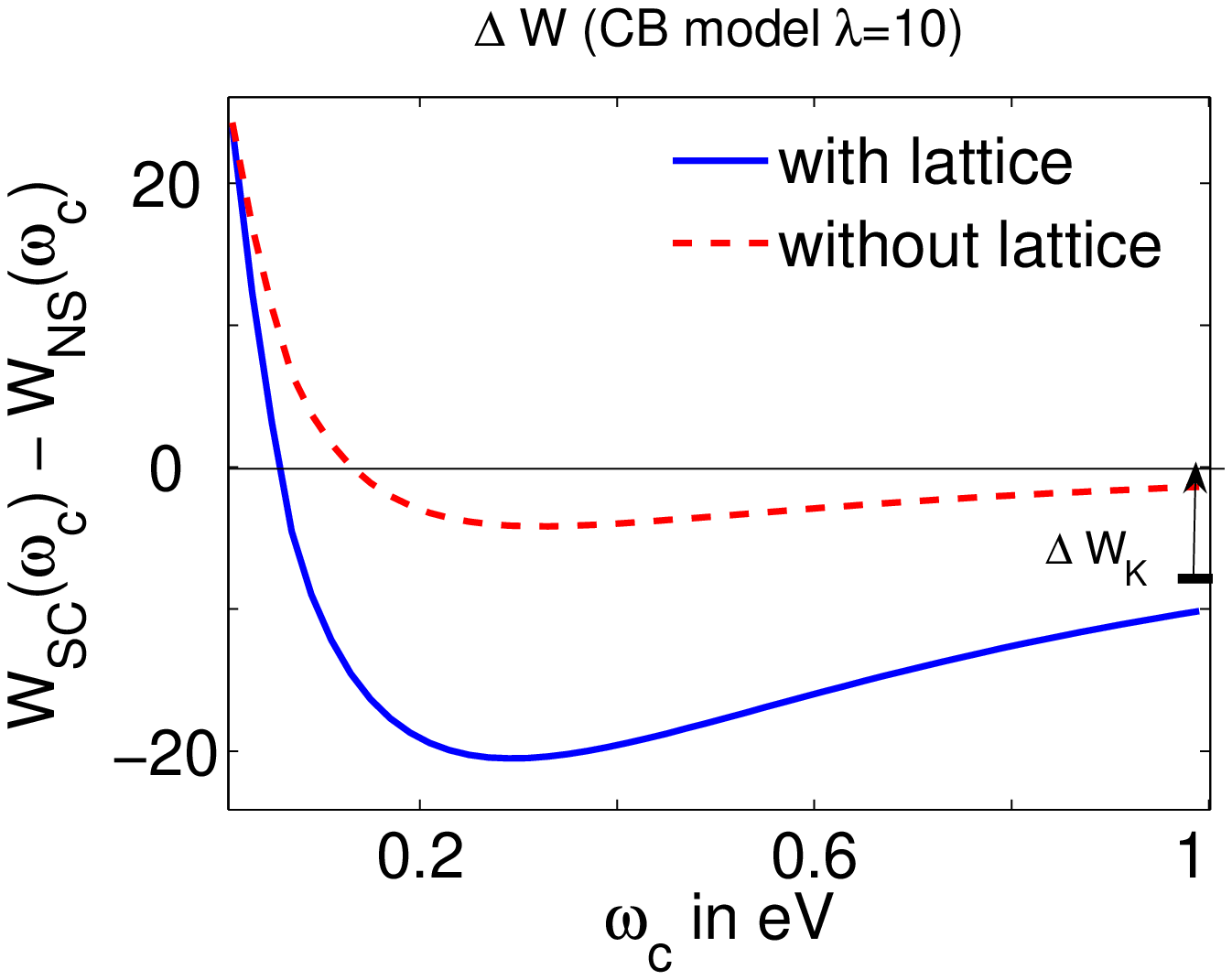}
\caption{\label{fig:CM_optdiff} $\Delta W$ (in meV)  for
$\lambda=1$(top) and $\lambda=10$(bottom). We used $\omega_{sf} =
26\,meV/\lambda$ and $\Delta = 30 meV$. The zero crossing is not
affected because we keep $\lambda \omega_{sf}$ constant. The
notable difference is the widening of the dip at a larger $\lambda$.}
\end{figure}

\begin{figure}[htp]
\includegraphics*[width=.9\linewidth]{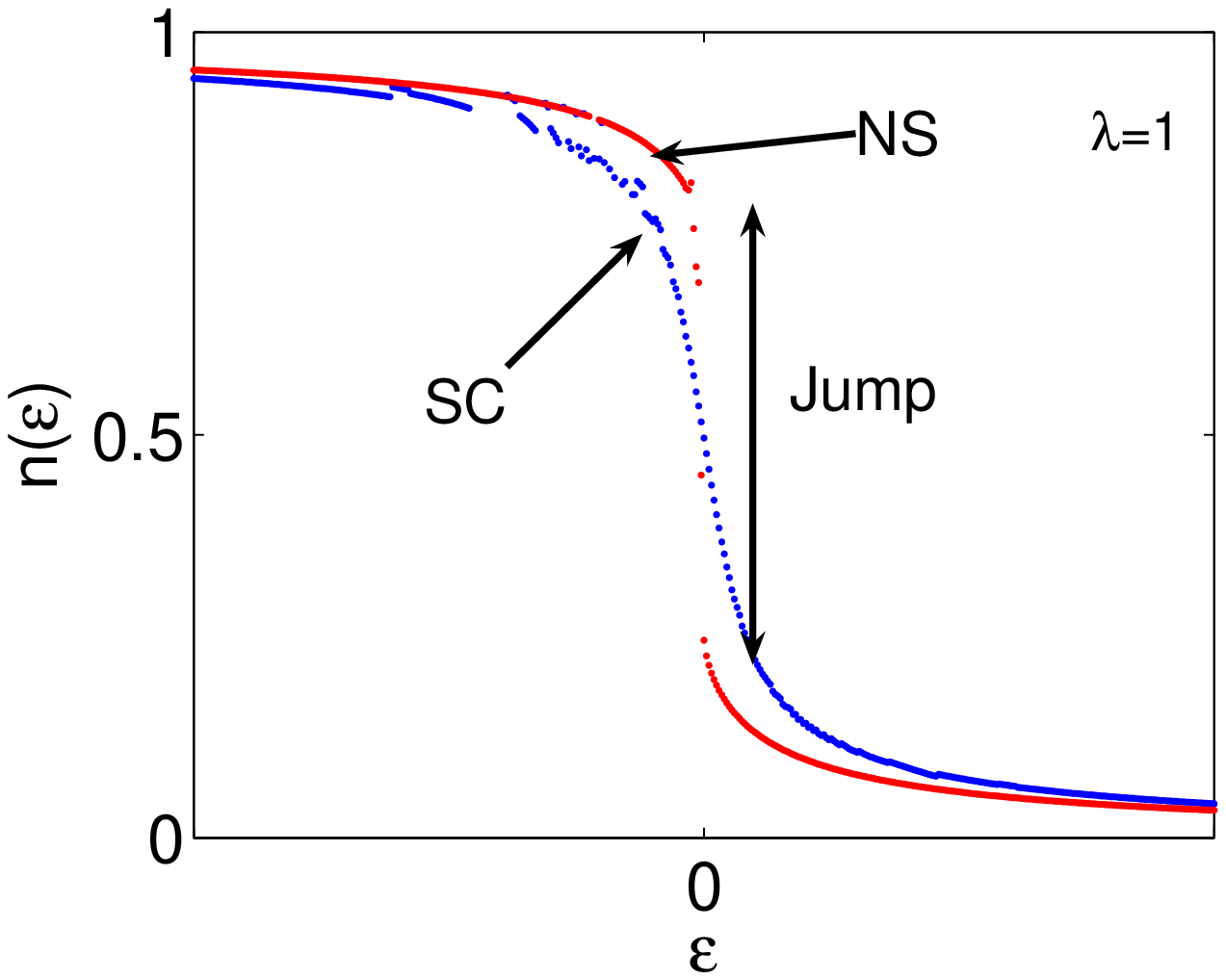}
\hfill
\includegraphics*[width=.9\linewidth]{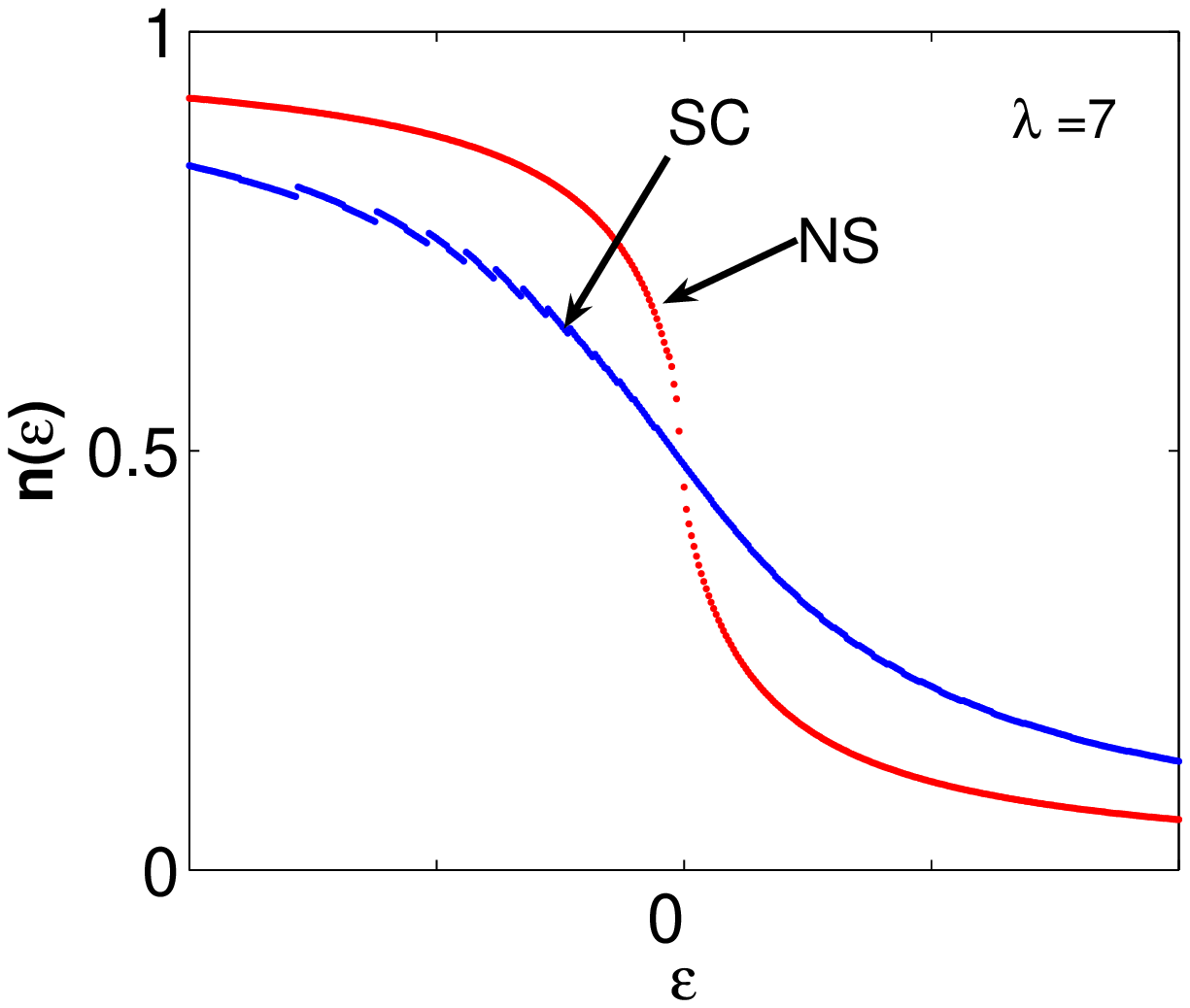}
\caption{\label{fig:CM_Dist fns}Distribution functions $n (\epsilon)$  for
CB model for $\lambda =1$ and $\lambda =7$ and a constant $\omega_{sf} =
26\,meV$. We set $\Delta = 30 meV$. For smaller $\lambda$ (top),
quasiparticles near the FS are well defined as indicated by the
well pronounced jump in $n (\epsilon)$.  For $\lambda =7$, $n (\epsilon)$
is rather smooth implying that a coherence is almost lost. Some irregularities
is the SCS distribution function are due to finite sampling in the
frequency domain. The irregularities disappear when finer mesh for
frequencies is chosen.}
\end{figure}

\begin{figure}[htp]
\includegraphics*[width=.7\linewidth]{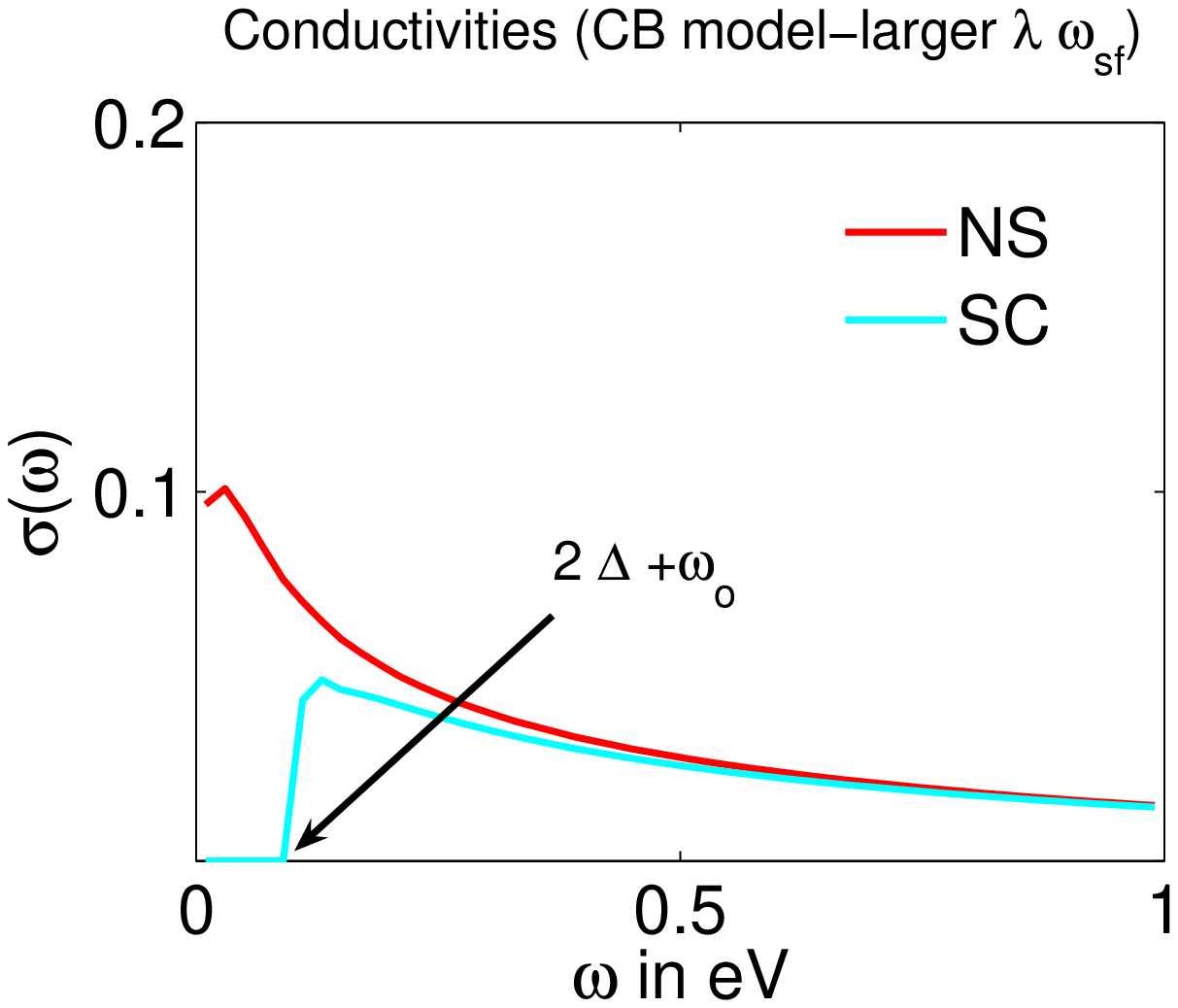}
\hfill
\includegraphics*[width=.7\linewidth]{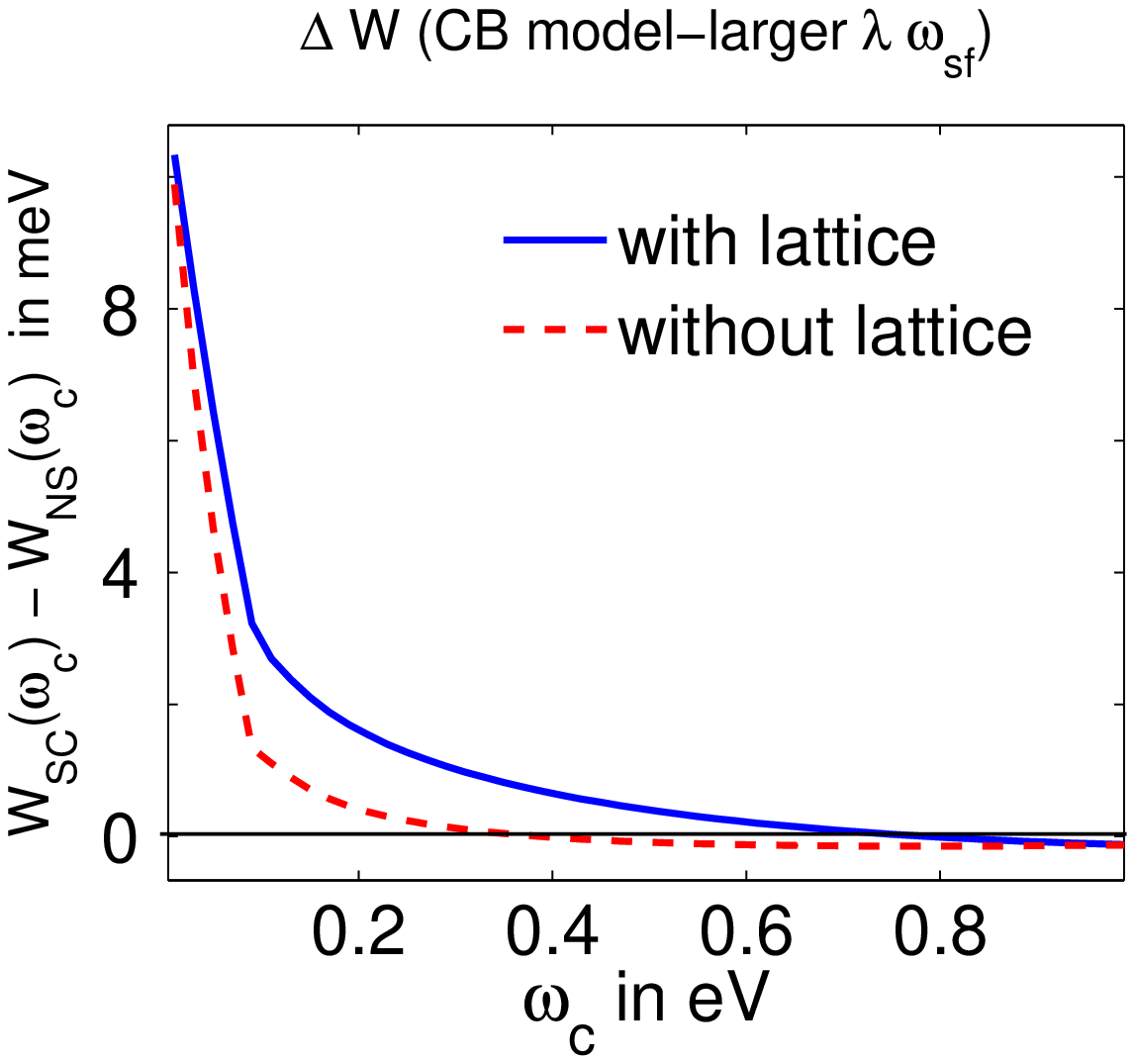}
\caption{\label{fig:CM_cond_L10_comp}Top -- conductivity at a larger
value of $\omega_{sf} \lambda$ ($\omega_{sf}=26\,meV$,$\lambda=7$)
consistent with the one used in Ref.\protect\onlinecite{bib:sum
rule mike_chu}). Bottom -- $\Delta W$ with and without lattice.
 Observe that the frequency of
zero crossing of $\Delta W$ enhances compared to the case of a smaller
 $\lambda \omega_{sf}$ and becomes  comparable to the bandwidth.
 At energies smaller than the bandwidth, $\Delta W >0$, as
in the Norman- P\'{e}pin model.}
\end{figure}

\begin{figure}[htp]
\includegraphics*[width=.7\linewidth]{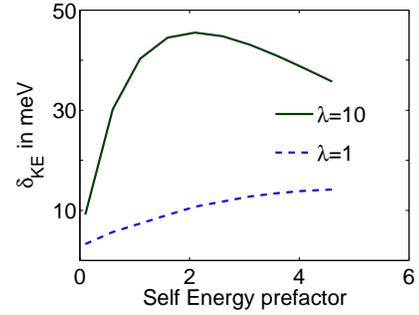}
\caption{\label{fig:CM_KE} Kinetic energy difference between the
SCS and the NS, $\delta_{KE}$ We set $\lambda$ to be either
$\lambda=1$ or $\lambda=10$ and varied $\omega_{sf}$  thus
changing the overall prefactor in the self-energy. At weak
coupling ($\lambda =1$) the behavior is BCS-like -- $\delta_{KE}$
is positive and increases with the overall factor in the self-energy.
 At strong coupling ($\lambda =7$), $\delta_{KE}$ shows a reverse trend at
larger $\omega_{sf}$.}
\end{figure}

We performed the same calculations of conductivities and optical
integrals as in the previous three cases. The results are
summarized in Figs.  \ref{fig:CM_cond} -
\ref{fig:CM_cond_L10_comp}.
 Fig \ref{fig:CM_cond} shows
conductivities in the NS and the SCS for two couplings $\lambda=1$
and $\lambda=10$ (keeping $\lambda \omega_{sf}$ constant). Other
parameters $Z_o$ and $\omega_o$ are calculated according to the
discussion after Eq \ref{eq:chi}. for $\omega_{sf}=26\,meV$,
$\lambda=1$, we find $\omega_o=40\,meV$, $Z_o=0.77$. And for
$\omega_{sf}=2.6\,meV$, $\lambda=10$, we find
$\omega_o=13.5\,meV$, $Z_o=1.22$. Note that the conductivity in
the SCS starts at $2\Delta + \omega_o$ (i.e. the resonance energy
shows up in the optical gap), where as in the BCSI case it would
have always begun from $2\Delta$. In Fig
\ref{fig:CM_Kubosums_mode40} we plot the Kubo sums $W_K$ vs
coupling $\lambda$. We see that for all $\lambda$, $W_K$ in the NS
stays larger than in the  SCS. Fig \ref{fig:CM_OS} shows the
cutoff dependence
 of the optical integrals $W (\w_c)$ for $\lambda =1$
 separately in the NS and the SCS.
We again see that only about $73\%$ of the Kubo sum is recovered up to
 the bandwidth of  $1\,eV$
indicating that there is a significant amount left to recover
beyond this energy scale. Fig \ref{fig:CM_optdiff} shows $\Delta
W$ for the two different couplings. We see that, for both
$\lambda$'s, there is only one zero-crossing for the $\Delta W$
curve, and $\Delta W$ is negative at larger frequencies. The only
difference  between the two plots is that for larger coupling the
dip in $\Delta W$ gets `shallower'. Observe also that the solid
line in  Fig. \ref{fig:CM_optdiff} is rather far away from the
dashed line at $\omega_c > 1 meV$, which indicates that, although
$\Delta W (\w_c)$ in this region has some dependence on $\w_c$,
still the largest part of  $\Delta W (\w_c)$ is $\Delta W_K$,
while the contribution from $\Delta f (\omega_c)$ is smaller.

The negative sign of $\Delta W (\w_c)$ above a relatively small
$\omega_c \sim 0.1-0.2 eV$ implies that the `compensating' effect
from the fermionic self-energy on $\Delta W$ is not strong enough
to overshadow the decrease of the optical integral in the SCS due
to gap opening. In other words,the CB model displays the same
behavior as  BCSI, EB, and modified MFLI models. It is interesting
that this holds despite the fact that for large $\lambda$
 CB model displays the physics one apparently needs to reverse the
sign of $\Delta W_K$ -- the absence of the quasiparticle peak in
the NS and  its emergence in the SCS accompanied by the dip and
the hump at larger energies. The absence of coherent quasiparticle
in the NS at large $\lambda$ is also apparent form Fig
\ref{fig:CM_Dist fns} where we show the normal state distribution
functions for two different $\lambda$. For large $\lambda$ the
jump (which indicates the presence of quasiparticles) virtually
disappears.

On a more careful look, we found that indifference of $\delta W
(\w_c)$ to the increase of $\lambda$ is merely the consequence of
the fact that above we kept $\lambda \omega_{sf}$ constant.
Indeed, at small frequencies, fermionic self-energy in the NS is
$\Sigma' = \lambda \omega$, $\Sigma " = \lambda^2
\omega^2/(\lambda \omega_{sf})$, and both $\Sigma'$ and $\Sigma''$
increase with $\lambda$ if we keep $\lambda \omega_{sf}$ constant.
But at frequencies larger than $\omega_{sf}$, which we actually
probe by $\Delta W (\w_c)$, the self-energy essentially depends
only on $\lambda \omega_{sf}$, and increasing $\lambda$ but
keeping   $\lambda \omega_{sf}$ constant does not bring us closer
to the physics associated with the recovery of electron coherence
in the SCS. To detect this physics, we need to see how things
evolve when we increase $\lambda \omega_{sf}$ above the scale of
$\Delta$ , i.e., consider a truly strong coupling when not only
$\lambda \gg 1$ but also
 the normal state $\Sigma_{NS} (\omega \geq \Delta) >>  \Delta$.

To address this issue, we took a larger $\lambda$ for the same
$\omega_{sf}$ and re-did the calculation of the conductivities and
optical integrals. The results for $\sigma (\omega)$  and $\Delta
W(\w_c)$ are presented in Fig. \ref{fig:CM_cond_L10_comp}. We
found the same behavior as before, i.e., $\Delta W_K$ is negative.
But we also found that the larger is the overall scale for the
self-energy, the larger is a frequency of zero-crossing of $\Delta
W(\w_c)$. In particular, for the same $\lambda$ and $\omega_{sf}$
that were used in Ref. \onlinecite{bib:sum rule mike_chu} to fit
the NS conductivity data, the zero crossing  is at $\sim 0.8\,eV$
which is quite close to the bandwidth. This implies that at a
truly strong coupling the frequency at which $\Delta W (\w_c)$
changes sign can well be larger than the bandwidth of $1 eV$ in
which case $\Delta W$ integrated up to the bandwidth does indeed
remain positive. Such  behavior would be consistent with
Refs.\onlinecite{bib:molegraaf, bib:optical int expt}. we also
see from  Fig. \ref{fig:CM_cond_L10_comp} that $\Delta W_K$
becomes small at a truly strong coupling, and over a wide range of
frequencies the behavior of $\Delta W (\w_c)$ is predominantly
governed by $\Delta f (\w_c)$, i.e. by the cut-off
term.~\cite{comm_pl} The implication is that, to first
approximation, $\Delta W_K$ can be neglected and positive $\Delta
W (w_c)$ integrated to a frequency where it is still positive is
almost compensated by the integral over larger frequencies. This
again would be consistent with the experimental data in Refs.
\onlinecite{bib:molegraaf, bib:optical int expt}.

It is also instructive to understand the interplay between the
behavior of $\Delta W(\w_c)$ and the behavior of the difference of
the kinetic energy between the SCS and the NS, $\delta_{KE}$. We
computed the kinetic energy as a function of $\lambda \omega_{sf}$
and present the results in Fig. \ref{fig:CM_KE} for $\lambda=1$
and $10$. For a relatively weak $\lambda=1$ the behavior is
clearly BCS like- $\delta_{KE} >0$ and increases with increasing
$\lambda \omega_{sf}$. However, at large $\lambda =10$, we see
that the kinetic energy begin decreasing  at large $\lambda
\omega_{sf}$ and eventually changes sign. The behavior of
$\delta_{KE}$ at a truly strong coupling is consistent with
earlier calculation of the kinetic energy for Ornstein-Zernike
form of the spin susceptibility~\cite{bib:haslinger}.

We clearly see that the increase of the zero crossing frequency of
$\Delta W (\w_c)$ at a truly strong coupling is correlated with
the non-BCS behavior of $\delta_{KE}$. At the same time, the
behavior of $\delta W (\w_c)$ is obviously not driven by the
kinetic energy as eventually $\delta W (\w_c)$ changes sign and
become negative. Rather, the increase in the frequency range where
$\Delta W (\w_c)$ remains positive and non-BCS behavior of
$\delta_{KE}$ are two indications of the same effect that fermions
are incoherent in the NS but acquire coherence in the SCS.

%%%%%%%%%%%%%%%%%%%%%%%%%%%%%%%%%%%%%%%%%%%%%%%%%%%%%%%%%%%%%%%%%%%%%%
%%%%%%%%%%%%%%%%%%%%%%%%%%%%%%%%%%%%%%%%%%%%%%%%%%%%%%%%%%%%%%%%%%%%%%
\section{Conclusion}

In this work we analyzed the behavior of optical integrals
$W(\w_c) \propto \int_o^{\w_c} \sigma (\omega) d \omega$ and  Kubo
sum rules in the normal and superconducting states of interacting
fermionic systems on a lattice. Our key goal was to understand
what sets the sign of $\Delta W_K = \Delta W(\infty)$ between the
normal and superconducting states and what is the behavior of
$W(\w_c)$ and $\Delta W(\w_c)$ at finite $\w_c$. In a weak
coupling BCS superconductor, $\Delta W (\w_c)$ is positive at $\w_c
<2\Delta$ due to a contribution from superfluid density, but
becomes negative at larger $\w_c$, and approach a negative value
of $\Delta W_K$.  Our study was motivated by fascinating optical
experiments on the
cuprates~\cite{bib:basov,bib:molegraaf,bib:boris,bib:optical int
expt}. In overdoped cuprates, there is clear
indication~\cite{bib:nicole} that $\Delta W (\w_c)$ becomes
negative above a few $\Delta$, consistent with BCS behavior. In
underdoped cuprates, two groups
argued\cite{bib:molegraaf,bib:optical int expt}  that $\Delta W$
integrated up to the bandwidth remains positive, while the other
group argued~\cite{bib:boris} that it is negative.

The reasoning why $\Delta W_K$ may potentially change sign at
strong coupling involves the correlation between $-W_K$ and the
kinetic energy. In the BCS limit, kinetic energy obviously
increases in a SCS because of gap opening, hence $-W_K$ increases,
and $\Delta W_K$ is negative. At strong coupling, there is a
counter effect -- fermions become more mobile in a SCS due to a
smaller self-energy.

We considered four models: a BCS model with impurities, a model of
fermions interacting with an Einstein boson, a phenomenological
MFL model with impurities, and a model of fermions interacting
with collective spin fluctuations. In all cases, we found that
$\Delta W_K$ is negative, but how it evolves with $\w_c$  and how
much of the sum rule is recovered by integrating up to the
bandwidth depends on the model.

The result most relevant to the experiments on the cuprates is
obtained for the spin fluctuation model. We found that at strong
coupling, the zero-crossing of $\delta W (\w_c)$ occurs at a
frequency which increases with the coupling strength and may
become larger than the bandwidth at a truly strong coupling.
Still, at even larger frequencies, $\Delta W (\w_c)$ is negative.

\section*{Acknowledgements}
We would like to thank M. Norman, Tom Timusk, Dmitri Basov, Chris
Homes, Nicole Bontemps, Andres Santander-Syro, Ricardo Lobo, Dirk
van der Marel, A. Boris, E. van Heumen, A. B. Kuzmenko, L.
Benfato, and F. Marsiglio  for many discussions concerning the
infrared conductivity and optical integrals and  thank A. Boris,
 E. van Heumen, J. Hirsch, and F. Marsiglio
for the comments on the manuscript.   The work was
supported by  NSF-DMR 0906953.


\begin{thebibliography}{5}
\bibitem{comm_pl} In this respect, our results are consistent with
the analysis of $\Delta W (\w_c)$ in a system without a lattice
(Ref.\protect\cite{bib:artem}). The authors of that work  also
found that the frequency of zero-crossing of $\Delta W (\w_c)$
increases with the coupling strength.
\end{thebibliography}
\end{document}